\documentclass[fleqn,usenatbib]{mnras}

\usepackage{newtxtext,newtxmath}

\usepackage[T1]{fontenc}

\usepackage{graphicx}	
\usepackage{amsmath}	
\usepackage{multicol} 
\usepackage{multirow} 
\usepackage[table]{xcolor}
\usepackage{caption} 
\usepackage{subcaption} 
\usepackage{orcidlink} 

\newcommand{\pcm}{\,cm$^{-2}$}	
\newcommand{\kmps}{\,km\,s$^{-1}$} 
\newcommand{\NH}{$N_\mathrm{H}$} 
\newcommand{\flux}{\,erg\,cm$^{-2}$\,s$^{-1}$} 
\newcommand{\Rg}{$R_\mathrm{g}$} 
\newcommand{\xmm}{\hbox{\textit{XMM-Newton}}} 
\newcommand{\nustar}{\hbox{\textit{NuSTAR}}}

\title[The hard X-ray variability of NGC 7582]{A hard look at the X-ray spectral variability of NGC 7582}

\author[M. Lefkir, E. Kammoun, ...]{
Mehdy Lefkir\,\orcidlink{0000-0002-6972-2429},$^{1,2}$\thanks{E-mail: \href{mailto:ml556@leicester.ac.uk}{ml556@leicester.ac.uk}, \href{mailto:mehdylefkir.work@gmail.com}{mehdylefkir.work@gmail.com}
}\,Elias Kammoun\,\orcidlink{0000-0002-0273-218X},$^{2,3}$ Didier Barret\,\orcidlink{0000-0002-0393-9190},$^{2}$ Peter Boorman\,\orcidlink{0000-0001-9379-4716},$^{4,5,6}$ Gabriele Matzeu\,\orcidlink{0000-0003-1994-5322},$^{7,8}$ \newauthor Jon M. Miller\,\orcidlink{0000-0003-2869-7682},$^{9}$
Emanuele Nardini\,\orcidlink{0000-0001-9226-8992},$^{3}$ Abderahmen Zoghbi\,\orcidlink{0000-0002-0572-9613}$^{10,11,12}$
\\
$^{1}$Department of Physics and Astronomy, University of Leicester, Leicester LE1 7RH, UK \\
$^{2}$IRAP, Université de Toulouse, CNRS, UPS, CNES 9, Avenue du Colonel Roche, BP 44346, F-31028, Toulouse Cedex 4, France\\
$^{3}$INAF – Osservatorio Astrofisico di Arcetri, Largo Enrico Fermi 5, I-50125 Firenze, Italy\\
$^{4}$Cahill Center for Astrophysics, California Institute of Technology, 1216 East California Boulevard, Pasadena, CA 91125, USA\\
$^{5}$ Astronomical Institute of the Czech Academy of Sciences, Boční II 1401/1A, 14100 Praha 4, Czech Republic\\
$^{6}$School of Physics \& Astronomy, University of Southampton, Highfield, Southampton SO17 1BJ, UK\\
$^{7}$Department of Physics and Astronomy (DIFA), University of Bologna, Via Gobetti 93/2, I-40129 Bologna, Italy\\
$^{8}$INAF – Osservatorio di Astrofisica e Scienza dello Spazio di Bologna, Via Gobetti 93/3, I-40129 Bologna, Italy\\
$^{9}$Department of Astronomy, University of Michigan, 1085 South University Avenue, Ann Arbor, MI 48109, USA\\
$^{10}$Department of Astronomy, University of Maryland, College Park, MD 20742\\
$^{11}$HEASARC, Code 6601, NASA/GSFC, Greenbelt, MD 20771\\
$^{12}$CRESST II, NASA Goddard Space Flight Center, Greenbelt, MD 20771\\
}

\date{Accepted XXX. Received YYY; in original form ZZZ}

\pubyear{2022}

\begin{document}
\label{firstpage}
\pagerange{\pageref{firstpage}--\pageref{lastpage}}
\maketitle

\begin{abstract}
 NGC~7582 ($z = 0.005264$; $D=22.5$~Mpc) is a highly variable, changing-look AGN. In this work, we explore the X-ray properties of this source using \xmm\ and \nustar\ archival observations in the 3-40 keV range, from 2001 to 2016. NGC~7582 exhibits a long-term variability between observations but also a short-term variability in two observations that has not been studied before. To study the variability, we perform a time-resolved spectral analysis using a phenomenological model and a physically-motivated model (\texttt{uxclumpy}). The spectral fitting is achieved using a nested sampling Monte Carlo method. \texttt{uxclumpy} enables testing various geometries of the absorber that may fit AGN spectra. We find that the best model is composed of a fully covering clumpy absorber. From this geometry, we estimate the velocity, size and distance of the clumps. The column density of the absorber in the line of sight varies from Compton-thin to Compton-thick between observations. Variability over the timescale of a few tens of kilo-seconds is also observed within two observations. The obscuring clouds are consistent with being located at a distance not larger than $0.6$~pc, moving with a transverse velocity exceeding $\sim 700$~\kmps. We could put only a lower limit on the size of the obscuring cloud being larger than $10^{13}~$cm. Given the sparsity of the observations, and the limited exposure time per observation available, we cannot determine the exact structure of the obscuring clouds. The results are broadly consistent with comet-like obscuring clouds or spherical clouds with a non-uniform density profile. 
\end{abstract}

\begin{keywords}
galaxies: active -- galaxies: Seyfert -- X-rays: galaxies -- accretion, accretion discs
\end{keywords}



\section{Introduction} 

Active Galactic Nuclei (AGN) are energetic sources exhibiting a high bolometric luminosity, located at galactic centres. They are powered by the accretion of matter onto a supermassive black hole with $M_\mathrm{BH}\gtrsim 10^{6}~\mathrm{M}_\odot$ via an inspiraling disc, known as the accretion disc. This disc is thought to be typically optically thick and geometrically thin \citep{1973A&A....24..337S}, emitting thermal radiation in the optical and ultraviolet range. In addition, a strong X-ray emission is observed in AGN and attributed to the up-scattering of disc photons by electrons in a hot plasma called the corona \citep{1993ApJ...413..507H}.

Using optical/UV spectra, active galaxies can be classified as: Type-1 that show narrow (Full Width at Half Maximum, FWHM, less than $1000$\kmps) and broad emission lines ($\rm FWHM > 1000$\kmps) and Type-2 that show narrow emission lines only. Optical lines are Doppler-broadened closer to the black hole in the eponymous region, Broad Line Region (BLR), in opposition to the Narrow Line Region (NLR) which is thought to be located farther from the black hole, where narrow lines are produced. The BLR is typically located at a distance less than one parsec \citep[see ][]{2005ApJ...629...61K} whereas the NLR can extend from dozens to hundreds of parsecs \citep{2012MNRAS.420..526M}.

The unified model of AGN \citep{1985ApJ...297..621A,1993ARA&A..31..473A,1995PASP..107..803U} explained the variety in AGN by the difference in viewing angle at which we observe the source. In this model, Type-1 sources are observed face-on providing a direct look at the central engine. Instead, Type-2 sources are observed with a higher inclination where obscuration of the central engine and BLR clouds by the parsec-scale dusty torus becomes more important. X-ray observations can be used to estimate the equivalent hydrogen column density ($N_{\rm H}$) of the obscurer which can be classified as Compton-thin or Compton-thick if $N_{\rm H}$ is smaller or larger than $\sim 1.5\times 10^{24}\,$\pcm, respectively. However, many sources, can change from Type-1 to Type-2 and vice versa between observations, associated with the appearance/vanishing  of broad optical lines \cite[see e.g., ][]{2015ApJ...800..144L}. \cite{2022arXiv221105132R} referred to these objects as Changing accretion-state (CS) AGN. Whereas, AGN changing from Compton-thick to Compton-thin and vice versa in X-ray observations are referred to as Changing-obscuration (CO) AGN. The proposed interpretations for CO-AGN transitions are the passage of obscuring material in the BLR across our line of sight or absorption by ionised material. In the case of CS-AGN, the obscuration of the source is ruled out in favour of disk instabilities or stellar tidal disruption events \citep[see e.g.,][]{2021ApJS..255....7R}. \\

In this work, we study NGC~7582, a nearby ($z=0.005264$, $D=22.5$~Mpc) Seyfert~2 galaxy with a changing-look behaviour. \cite{2006A&A...460..449W}  estimated the black hole mass to be $M_\mathrm{BH}={5.5^{+3.2}_{-2.3}\times10^7\,\mathrm{M}_\odot}$ by modelling the gas dynamics in the mid-infrared. This source has been observed in X-rays in several occasions since the late 70's. \cite{1978ApJ...223..788W} associated the X-ray source 2A 2315-428, detected with the \textit{Ariel 5} survey, with the galaxy NGC\,7582. These observations revealed the presence of a flux variability by a factor of 4 over 60 days. \cite{1982ApJ...256...92M} analysed three \textit{HEAO-1} observations. The source revealed short timescale variability with a flare where the count-rate doubled over a timescale of $1.5~$days. Using a \textit{Ginga} observation, \cite{1993MNRAS.265..412W} established the presence of a large intrinsic column density of absorbing material in the nuclear region. Later on, \cite{1999ApJ...519L.123A} reported on a broadening of the H$\alpha$ line, which involved a transition from Type-2 to Type-1, which labelled NGC~7582 as a changing-look AGN. \xmm\ observed the source for the first time in 2001, catching the source in a low-flux state. \cite{2005ApJ...625L..31D} and \cite{2006ApJ...645..928A} analysed this observation, and confirmed the detection of a clear iron line at $\sim 6.4$~keV. \cite{2007MNRAS.374..697B} mapped the nuclear region using \textit{Chandra} and the \textit{Hubble Space Telescope}. The authors report the presence of a dust lane in the optical, and concluded that soft X-ray emission (below $3$\,keV) corresponds to the region where the dust lane is thinner. \cite{2007AA...466..855P} analysed \xmm\ data from observations of 2001 and 2005. They proposed a model for the absorption that consists of a thin neutral absorber and an inner thick neutral absorber composed of clouds where X-ray photons are reflected via Compton scattering. \cite{2009ApJ...695..781B} analysed four \textit{Suzaku} and one \xmm\ observation, taking place in 2007, which were compared to the previous \xmm\ observations of 2001 and 2005. The authors observed an increase in the column density of the inner thick absorber during the \textit{Suzaku} campaign. They invoked a third, Compton-thin, absorber in their model, that is associated with the dust lanes outside of the torus observed with \textit{Chandra} and \textit{HST}. The authors ascribed the observed variability to absorbing material drifting in the BLR. Later on, \cite{2015ApJ...815...55R} studied two \nustar\ observations and one \textit{Swift} observation, taken in 2012. They modelled the absorption with a thick patchy torus and a thin full-covering absorber. They found the most plausible scenario  explaining the variability to be the obscuration of the nucleus. More recently, \cite{2017AA...600A.135B} analysed two \textit{Chandra} HETG observations performed in 2014, and a \textit{Suzaku} one performed in 2007. They found a constant reflection component associated with the inner edge of the clumpy torus. A highly ionised absorber was also detected in the \textit{Chandra} spectra.

In this work, we present a comprehensive analysis of all the archival \xmm\ and \nustar\ observations of NGC\,7582 with the aim of explaining the short- and long-term behaviour of the source. The paper is organised as follows. In Section~\ref{sec:observations}, we describe the observations analysed in this work and the corresponding data reduction. In Section~\ref{sec:SpectralAnalysis} we present the tools and models used for the X-ray spectral analysis. Our results are presented in Section~\ref{sec:Results} and discussed in Section~\ref{sec:Discussion}. Conclusions are drawn in Section~\ref{sec:Conclusion}.



\section{Observations and data reduction}
\label{sec:observations}
NGC~7582 was observed five times by \xmm\ between 2001 and 2018 and three times by  \nustar\ between 2012 and 2016, including a simultaneous observation between \xmm\ and \nustar\ in 2016. We use all of the archival \xmm\ and \nustar\ observations, except the \xmm\ observation of 2018. The source was off-axis by more than 10~arcmin, falling on a bad pixel column. The data from this observation could not be used as they suffered from calibration issues in energy and normalisation. We present in Table~\ref{tab:observations} the observations used in this work and the associated net exposure times.

\begin{table}
\centering
\caption{Observations used in this work and their associated exposure time after removing the bad time intervals. For \xmm\ the exposures are given for the EPIC instruments pn, MOS1 and MOS2, and for \nustar\ the net exposures are given for FPMA and FPMB. }
\begin{tabular}{llll}\hline\hline
Observation ID & Observatory & Date & Exposure (ks)\\\hline
0112310201  & \xmm    & 25 May 2001 & 17.5 / 22.4 / 22.4 \\
0204610101  & \xmm    & 29 Apr 2005 & 61.8 / 75.4 / 76.3 \\
0405380701  & \xmm    & 30 Apr 2007 & 1.9 / 8.6 / 8.6 \\
60061318002 & \nustar & 31 Aug 2012 & 16.5 / 16.4 \\
60061318004 & \nustar & 14 Sept 2012 & 14.6 / 14.6 \\
0782720301  & \xmm    & 28 Apr 2016 & 42 / 98.5 / 98.5 \\
60201003002 & \nustar & 28 Apr 2016 & 48.5 / 48.3 \\   
\hline\hline
\end{tabular}
\label{tab:observations}
\end{table}

\subsection{\textit{XMM-Newton}}

\textit{XMM-Newton} data from the EPIC instruments pn \citep{2001A&A...365L..18S} and MOS \citep{2001A&A...365L..27T} were reduced using the Science Analysis System  \citep[\texttt{SAS 20.0}; ][]{2004ASPC..314..759G}. We used the recent empirical calibration\footnote{\url{https://xmmweb.esac.esa.int/docs/documents/CAL-SRN-0388-1-4.pdf}} to improve the spectral agreement between \xmm\ and \textit{NuSTAR} above 3~keV. We filtered periods of high background flaring activity in the event lists with the recommended rate threshold of $0.4$ $\rm Count~s^{-1}$ for the EPIC-pn and $0.35$ $\rm Count~s^{-1}$ for the EPIC-MOS detectors. The source was extracted from a circular region of $30~''$ and the background was extracted from a circular source-free region of the same chip with a radius of $60~''$. Source and background light-curves were extracted in the soft energy band 0.5-3~keV and in the hard energy band 3-10~keV using the task \texttt{evselect}. A binning of 1~ks was chosen for all observations. \textit{XMM-Newton} source and background spectra are extracted with the task \texttt{evselect}. The Ancillary Response File (ARF) is generated with \texttt{arfgen} and the Redistribution Matrix File  (RMF) is produced with \texttt{rmfgen}. The spectral files are grouped with \texttt{specgroup} to have a minimum signal-to-noise ratio of five and ensuring that the width of the bins is not narrower than one-third of the energy resolution.

\begin{figure*}
    \centering
    \includegraphics[width=\textwidth]{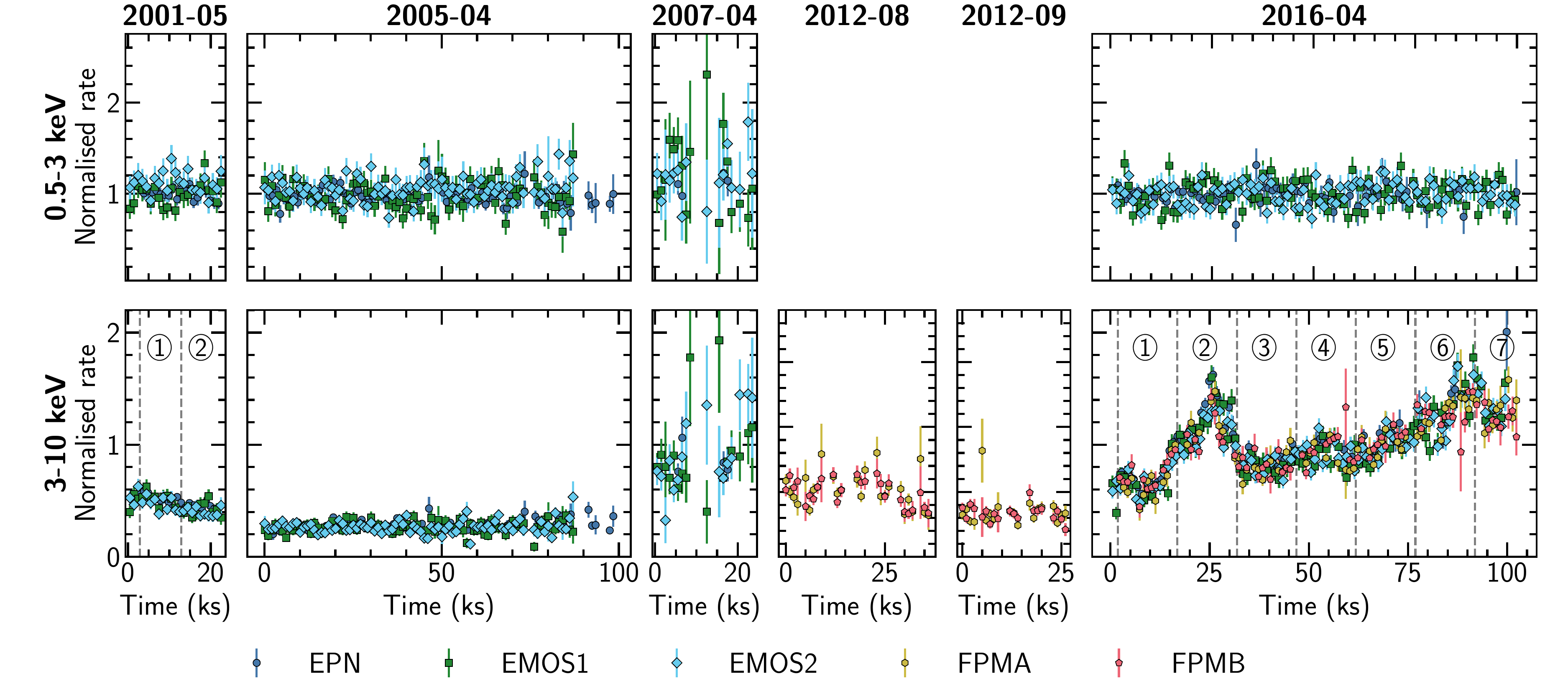}
    \caption{Light-curves of the \xmm\ and \nustar\ observations in the 0.5-3~keV (top panel) energy band and 3-10~keV (bottom panel) energy band for the EPIC-pn (dark blue circles), EPIC-MOS1 (green squares), EPIC-MOS2 (light blue diamonds), FPMA (yellow hexagons) and FPMB (magenta pentagons). The count-rates are normalised to the count-rates of the observation of 2016. The dashed vertical lines show the chosen slices for the extraction of spectra, the slices are indicated by circled numbers 
    \textcircled{1}, \textcircled{2}, \textcircled{3}, ...}
    \label{fig:all_lightcurve}
\end{figure*}

\subsection{\textit{NuSTAR}}

\textit{NuSTAR} \citep{2013ApJ...770..103H} data were reduced using the \texttt{NUSTARDAS~2.1.1} package part of \texttt{HEAsoft~6.29} and the calibration files of \texttt{CALDB 20210210}. The task \texttt{nupipeline} was used to produce calibrated data as well as filtered event lists. Source and background regions were selected as circular regions with a radius of 80\arcsec\ on both instruments FPMA and FPMB. Source and background light-curves were extracted for the 3-10~keV and 10-40~keV energy bands using the task \texttt{nuproducts}. We use \nustar\ data only up to 40~keV as they are background dominated above this energy. To keep consistency with the \xmm\ observations a temporal binning of 1~ks was chosen. Source and background spectra, ARF, and RMF, are extracted using the task \texttt{nuproducts} and grouped to at least 30 counts per energy bin.

\begin{figure}
    \centering
    \includegraphics[width=\columnwidth]{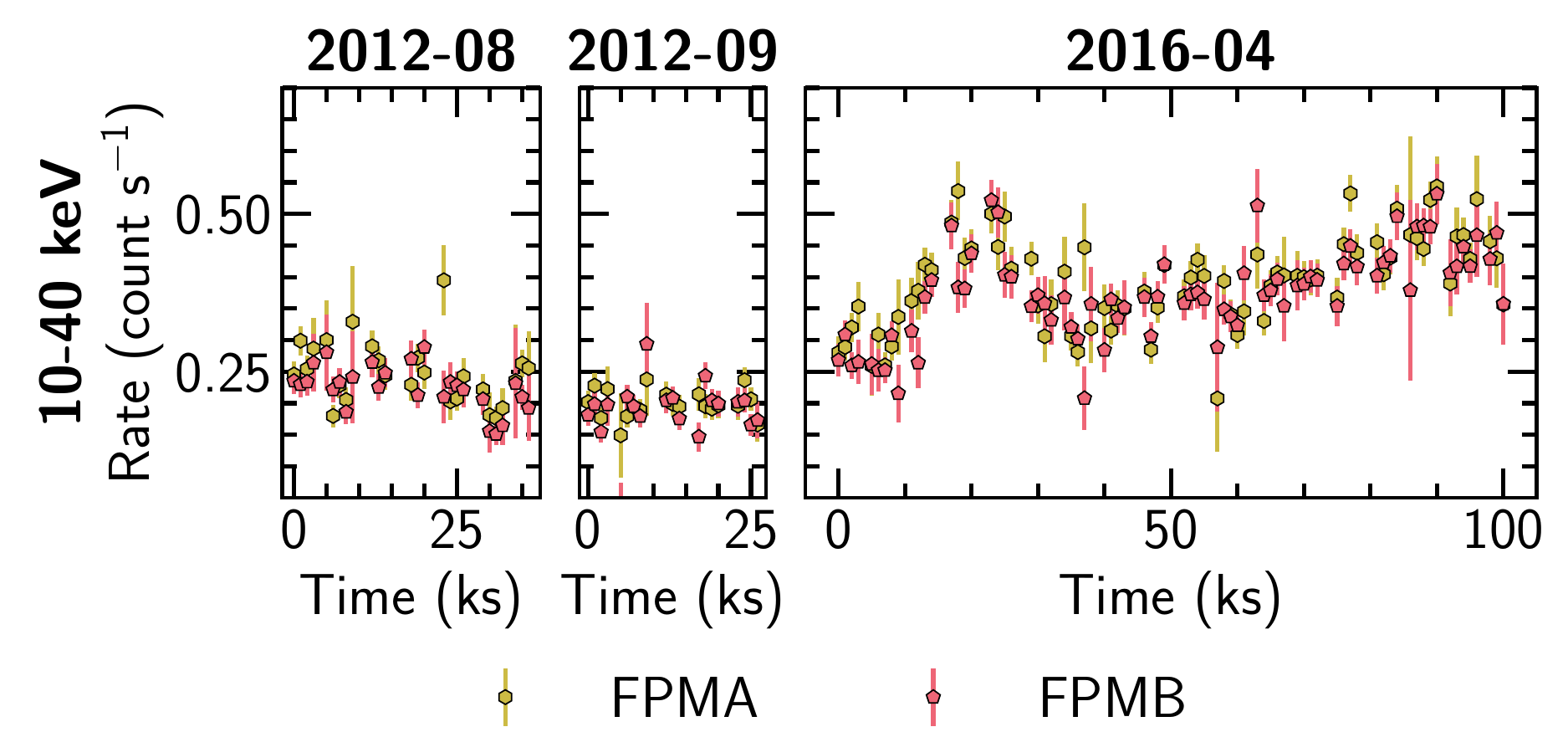}
    \caption{Light-curves  of \nustar\ observations in the 10-40~keV band for the detectors FPMA (yellow hexagons) and FPMB (magenta pentagons).}
    \label{fig:nustar_lightcurve}
\end{figure}

\begin{table}
\centering
\caption{Mean count-rate (Counts~$\mathrm{s}^{-1}$) of \xmm\ and \nustar\ observations in the 0.5-3~keV and 3-10~keV band.}

\begin{tabular}{lccccc}\hline\hline
& \multicolumn{5}{c}{Mean count-rate (Counts~$\mathrm{s}^{-1}$)}   \\ \hline
& \multicolumn{3}{c}{\begin{tabular}[c]{@{}c@{}}0.5-3 keV\\ 3-10 keV\end{tabular}}                                                  & \multicolumn{2}{c}{3-10 keV} \\ \hline
Date & pn  & MOS1  & MOS2      & FPMA          & FPMB         \\\hline
2001-05 & \begin{tabular}[c]{@{}l@{}}$0.282$\\ $0.276$\end{tabular} & \begin{tabular}[c]{@{}l@{}}$0.0792$\\ $0.0779$\end{tabular} & \begin{tabular}[c]{@{}l@{}}$0.0887$\\ $0.0828$\end{tabular} & - & -  \\\hline
2005-04 & \begin{tabular}[c]{@{}l@{}}$0.274$\\ $0.159$\end{tabular} & \begin{tabular}[c]{@{}l@{}}$0.0779$\\ $0.0428$\end{tabular} & \begin{tabular}[c]{@{}l@{}}$0.0828$\\ $0.0461$\end{tabular} & - & -  \\\hline
2007-04 & \begin{tabular}[c]{@{}l@{}}$0.319$\\ $0.462$\end{tabular} & \begin{tabular}[c]{@{}l@{}}$0.0946$\\ $0.134$\end{tabular}  & \begin{tabular}[c]{@{}l@{}}$0.931$\\ $0.140$\end{tabular}    & - & - \\ \hline
2012-08 & - & - & - & $0.193$ & $0.194$  \\\hline
2012-09 & - & - & - & $0.127$ & $0.132$  \\\hline
2016-04 & \begin{tabular}[c]{@{}l@{}}$0.285$\\ $0.521$\end{tabular} & \begin{tabular}[c]{@{}l@{}}$0.0784$\\ $0.154$\end{tabular}  & \begin{tabular}[c]{@{}l@{}}$0.0787$\\ $0.169$\end{tabular}  & $0.377$       & $0.386$ \\ \hline\hline
\end{tabular}
\label{tab:countrates}
\end{table}

\subsection{Short-term and long-term Variability}

The light-curves of the \xmm\ and \nustar\ observations are shown in Figure~\ref{fig:all_lightcurve}, normalised to the mean count-rates of the observation of 2016. 
The mean count-rates of the observations are given in Table~\ref{tab:countrates}. The light-curves in the 0.5-3\,keV range do not show any intra-/inter-observation variations. However, the variability is larger in the 3-10\,keV range.
The light-curve of the observation of 2001 appears to be decreasing towards the end of the observation. No variability is observed in the observation of 2005, which has the lowest count-rate among the analysed observations. For the observation of 2007, as the light-curve does not have sufficient data points, we cannot derive any conclusions on its variability. The two observations of 2012 do not present any strong short-term variability. The simultaneous observation of 2016 presents a strong variability with a flare appearing $\sim 15$\,ks after the start of the observation. Figure~\ref{fig:nustar_lightcurve} shows the light-curves of the \nustar\ observations in the 10-40~keV band. No strong variability is observed in 2012. The 10-40\,keV light-curve in 2016 shows similar variability to the one observed in the 3-10\,keV range.

In Table~\ref{tab:variability}, we compute the normalised excess variance  $\sigma_\mathrm{NXS}$ and  the fractional variability amplitude $F_\mathrm{var}$  following \cite{2003MNRAS.345.1271V}. In the 0.5-3~keV band, the mean square error of the measurement is larger than the sample variance, therefore no variability is detected. In the 3-10~keV band, the observation of 2016 presents a fractional variability amplitude always greater than $20\%$. The observations of 2012 also  present significant values of $F_\mathrm{var}$ hinting at a possible variability on a timescale longer than these short observations.

\begin{table*}
\centering
\caption{Normalised excess variance $\sigma_\mathrm{NXS}$ and fractional variability amplitude $F_\mathrm{var}$ of \xmm\ and \nustar\ observations for the EPIC-pn in the 0.5-3~keV and 3-10~keV band and for FPMA/FPMB in the 3-10~keV and 10-40~keV.}{

\begin{tabular}{lllllll}\hline\hline
& \multicolumn{2}{c}{\begin{tabular}[c]{@{}c@{}}pn\\0.5-3 keV\\ 3-10 keV\end{tabular}}                & \multicolumn{2}{c}{\begin{tabular}[c]{@{}c@{}}FPMA\\ 3-10 keV\\ 10-40 keV\end{tabular}}            & \multicolumn{2}{c}{\begin{tabular}[c]{@{}c@{}}FPMB\\ 3-10 keV\\ 10-40 keV\end{tabular}}                      \\ \hline
Date & $\sigma_\mathrm{NXS} ( 10^{-3})$  & $F_\mathrm{var} (\%)$  & $\sigma_\mathrm{NXS} ( 10^{-2})$ & $F_\mathrm{var} (\%)$ & $\sigma_\mathrm{NXS} ( 10^{-2})$  & $F_\mathrm{var} (\%)$   \\\hline

2001-05 & \begin{tabular}[c]{@{}l@{}}$-1.1\pm1.6$\\ $7.9\pm3.3$\end{tabular} & 
\begin{tabular}[c]{@{}l@{}}$-$\\ $8.7\pm1.8$\end{tabular} &    & &     &   \\ \hline

2005-07 & \begin{tabular}[c]{@{}l@{}}$-3.2\pm1.5$\\ $4.9\pm3.4$\end{tabular} & 
\begin{tabular}[c]{@{}l@{}}$-$\\ $7.0\pm2.4$\end{tabular} &    & &     &   \\ \hline

2007-04 & \begin{tabular}[c]{@{}l@{}}$-17\pm18$\\ $5.0\pm18$\end{tabular}   & 
\begin{tabular}[c]{@{}l@{}}$-$\\ $7.1\pm13$\end{tabular} &    & &     &   \\ \hline

2012-08 &   &   & \begin{tabular}[c]{@{}l@{}}$3.4\pm1.6$\\ $1.5\pm1.0$\end{tabular}  &
\begin{tabular}[c]{@{}l@{}}$18\pm4.4$\\ $12\pm4.2$\end{tabular} &
\begin{tabular}[c]{@{}l@{}}$1.2\pm1.1$\\ $-0.21\pm0.7$\end{tabular} & 
\begin{tabular}[c]{@{}l@{}}$11\pm5.0$\\ -\end{tabular}      \\ \hline

2012-09 &   &   & \begin{tabular}[c]{@{}l@{}}$7.2\pm2.8$\\ $-0.75\pm0.64$\end{tabular} & 
\begin{tabular}[c]{@{}l@{}}$27\pm5.3$\\ $-$\end{tabular}    & 
\begin{tabular}[c]{@{}l@{}}$0.09\pm1$\\ $5.1\pm1.6$\end{tabular}  & 
\begin{tabular}[c]{@{}l@{}}$2.9\pm17$\\$22.5\pm3.5$\end{tabular} \\ \hline

2016-04 & \begin{tabular}[c]{@{}l@{}}$-1.6\pm2$\\ $68\pm6$\end{tabular}  & 
\begin{tabular}[c]{@{}l@{}}$-$\\ $26\pm1$\end{tabular}  & 
\begin{tabular}[c]{@{}l@{}}$5.8\pm0.6$\\ $2.3\pm0.4$\end{tabular}  & 
\begin{tabular}[c]{@{}l@{}}$24\pm1.3$\\ $15\pm1.3$\end{tabular} & 
\begin{tabular}[c]{@{}l@{}}$4.1\pm0.5$\\ $2.3\pm0.4$\end{tabular}   &
\begin{tabular}[c]{@{}l@{}}$20\pm1.3$\\ $15\pm1.4$\end{tabular}\\\hline\hline 
\end{tabular}}

\label{tab:variability}
\end{table*}

To study the short-term spectral variability of this source, we decide to slice the observation of 2001 in two slices of 10~ks denoted by circled numbers \textcircled{1} and \textcircled{2}. The simultaneous observations of 2016 are sliced in seven slices of 15~ks indexed from \textcircled{1} to \textcircled{7}. The start of the EPIC-pn exposure serves as the origin of time for the intervals. The intervals are delimited by dashed vertical lines in the light-curves of Figure~\ref{fig:all_lightcurve}.



\section{Spectral analysis}
\label{sec:SpectralAnalysis}

In this work, the spectral analysis is performed with the X-ray Spectral Fitting Package \citep[XSPEC;][]{1996ASPC..101...17A} and its Python wrapper PyXspec. We adopt the cross-sections of \cite{1996ApJ...465..487V} and the chemical abundances reported in \cite{2000ApJ...542..914W}.

Cash statistic \citep[C-stat; ][]{1979ApJ...228..939C} is used to account for the Poisson distribution of photons. The fitting is performed using the Bayesian X-ray Analysis package \citep[BXA\footnote{\url{https://johannesbuchner.github.io/BXA/index.html}};][]{2014A&A...564A.125B} bridging XSPEC with nested sampling Monte Carlo algorithm implemented in UltraNest \citep{2016S&C....26..383B,2019PASP..131j8005B,2021JOSS....6.3001B}. We use nested sampling for its unsupervised global parameter exploration and  its naturally self-convergence criterion. We use wide log-uniform priors for normalisation and column density parameters and wide uniform priors for the other parameters. Using samples of the posterior distribution of the parameters, the median of the parameters is estimated. Uncertainties at $68\%$ confidence levels are computed using the 16th and 84th quantiles of the posterior distributions. 

Cross-calibration factors between instruments are included in all models. These factors are estimated by fitting the phenomenological model. We assess the goodness of fit through simulations with {\tt fakeit} in XSPEC using the best-fit value obtained from the fit. We simulate 1000 spectra for each instrument and rebin the spectra with a constant number of channels, 50 and 30, for \xmm\ and \nustar, respectively. The realisations are then plotted against the observed data. Due to the lack of variability below 3~keV, we decided to perform the spectral analysis only above 3~keV. We found a cross-calibration  mismatch between \xmm\ and \nustar\ below $\sim4$~keV, for that reason we fit \nustar\ data only above that energy.

\subsection{Phenomenological model}
\label{sec:phenommodel}

First, we fitted a phenomenological model to the spectra, which can be written in XSPEC parlance as follows,

\begin{equation}
\centering
    \texttt{Model =} \texttt{constant}\times \texttt{TBabs} \times ( \texttt{cutoffpl} \times \texttt{TBpcf} + \texttt{xillver} ).
    \label{eq:phenommodel}
\end{equation}

\noindent The first component \texttt{constant} is the cross-calibration factor between the different instruments. The second component \texttt{TBabs} \citep{2000ApJ...542..914W} models the absorption of X-ray photons in the line of sight by the gas and dust of our Galaxy with an equivalent hydrogen column density $N_\mathrm{H}=1.26 \times 10^{20}$~\pcm  \citep{2016A&A...594A.116H}. The primary X-ray emission from the hot corona is modelled using a power-law with an exponential cutoff (\texttt{cutoffpl}). We tried to let the high-energy cutoff free to vary but no constraints were obtained so it was frozen to 500~keV for all the fits.
We accounted for the absorption of the primary emission with \texttt{TBpcf} \citep{2000ApJ...542..914W} where the partial covering fraction $f_{\rm cov}$ was left free to vary. As their values converged to unity we fixed $f_{\rm cov}=1$, modelling a fully covering absorber, for the remainder of the analysis. The primary emission in Type-2 sources is thought to be reprocessed by distant neutral material leading to a reflection spectrum that we modelled using \texttt{xillver} \citep{2010ApJ...718..695G,2013ApJ...768..146G,2014ApJ...782...76G}.  \texttt{xillver} is a tabulated model, computed by solving the radiative transfer equation in a plane-parallel geometry for a slab of gas illuminated by an X-ray source. We assume the reflector as distant and neutral, so the ionization parameter $\log\xi$ is frozen to 0. As NGC~7582 is a Seyfert-2 we freeze the inclination of the line of sight with respect to the normal of the disk to $\theta=80^\circ$. The photon index and the high-energy cutoff of \texttt{xillver} are tied to the ones of the illuminating power law.

We added a scattered primary emission component in the form of a constant times a power law to account for the scattered emission observed for instance in \cite{1997ApJS..113...23T}. The constant was allowed to vary between 0 and 1 to model the scattered fraction, the parameters of the power law are tied to the ones of the primary power law. When fitting the phenomenological model with this component for the first slice of the observations of 2001 and 2016 and the observation of 2005 we find values of the scattered fraction less than $2\%$. These results are consistent with the relation between the column density and the scattered fraction observed in \cite{2021MNRAS.504..428G}. In fact, this component is supposed to be more important at soft X-rays, below $\sim 2-3\,\rm keV$, which are not analysed in this work. Thus, we do not include this component in the model as it does not improve the quality of the fit substantially.

\subsection{Physically-motivated model}

We also use the physical Unified X-ray Clumpy model \citep[{\tt uxclumpy}\footnote{\url{https://github.com/JohannesBuchner/xars/blob/master/doc/uxclumpy.rst}}; ][]{2019A&A...629A..16B}.  This model takes into account Compton scattering, photo-electric absorption and line fluorescence effects due to the absorber. \cite{2019A&A...629A..16B} produced this clumpy model assuming that the total number of clouds is constant but the observed distribution depends on the geometry of the absorber and the line of sight. The distribution of the angular size of the clumps is computed to reproduce eclipse events in AGN.

Two geometrical parameters are used: {\tt TORsigma}, i.e. the vertical extent of the cloud population, the smaller this value the more concentrated the clouds in the disc plane; and {\tt CTKcover}, i.e. the covering factor of the inner ring of clouds, the higher this value the larger the inner clouds. 
The inclination angle of the line of sight with respect to the normal to the disc is frozen to $80^\circ$ as in the phenomenological model.

In XSPEC, the model can be written as follows,
\begin{multline}
   \texttt{Model =} \texttt{constant} \times \texttt{TBabs}\times (\texttt{atable[uxclumpy-cutoff]} \\+\texttt{constant}\times\texttt{atable[uxclumpy-cutoff-omni]}).
    \label{eq:physicalmodel}
\end{multline}

\noindent As mentioned previously, the first \texttt{constant} component is the cross-calibration factor estimated with the phenomenological model and \texttt{TBabs} models the absorption of X-rays by gas and dust of our Galaxy. The table \texttt{uxclumpy-cutoff} models the spectrum with the reflected and transmitted components as well as fluorescent lines. The table \texttt{uxclumpy-cutoff-omni} models the incident spectrum of the X-ray corona with the power law. This component models unabsorbed emission of the corona, hence it is supposed to model a warm re-emission. The parameters of the \texttt{uxclumpy-cutoff-omni} component are tied to the ones of \texttt{uxclumpy-cutoff}, as they must represent the same geometry. A factor ranging between $10^{-5}$ and $0.1$ is added for the warm emission model \texttt{uxclumpy-cutoff-omni}, this limits the contribution of the warm emission component up to $10\%$ of the transmitted and reflected component.

\section{Results}
\label{sec:Results}

\subsection{Phenomenological model}
\begin{figure*}
    \centering
    \includegraphics[width=\textwidth]{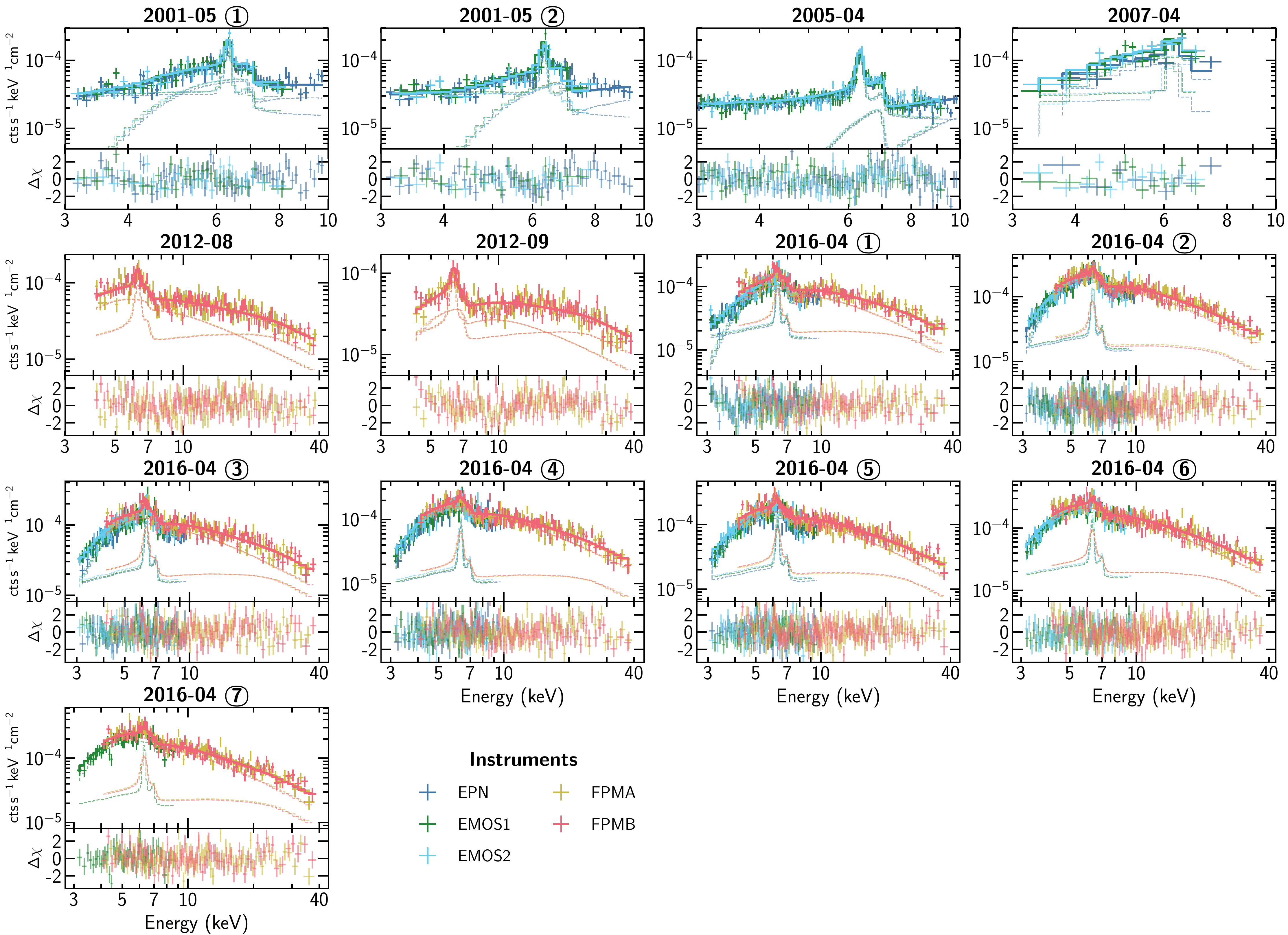}
    \caption{Phenomenological model fitted on all epochs with the spectra (upper panels) and residuals (lower panels). The instruments are plotted as follows: EPIC-pn in dark blue, EPIC-MOS1 in green, EPIC-MOS2 in light blue, FPMA in yellow and FPMB in magenta. The additive individual components of the model are depicted in coloured dashed lines. The circled numbers \textcircled{1}, \textcircled{2}, ...,  \textcircled{7} indicate the slices corresponding to the given observation.}
    \label{fig:phenom_spectra}
\end{figure*}

\begin{figure}
    \centering
    \includegraphics[width=\columnwidth]{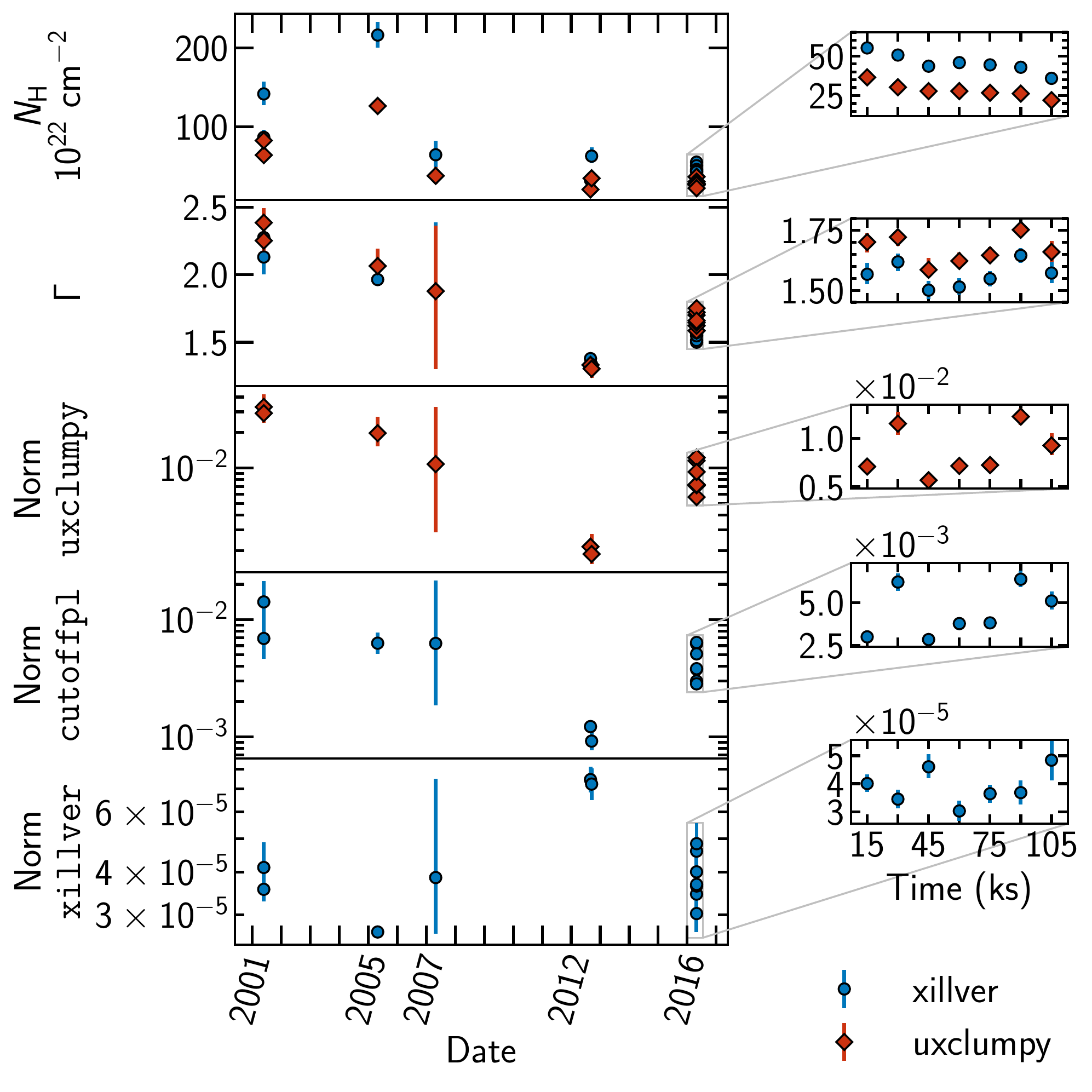}
    \caption{Best fit values of the best physical model (see Section \ref{sec:physicalmodel}) (red diamonds) and phenomenological model (blue circles) in all epochs, given within $68\%$-confidence levels. Panels from top to bottom: equivalent hydrogen column density, photon index, normalisation of \texttt{uxclumpy}, normalisation of \texttt{cutoffpl} and normalisation of \texttt{xillver}.}
    \label{fig:evolution}
\end{figure}

We fit the phenomenological model presented in Section~\ref{sec:phenommodel}, finding good fits in all epochs. The best-fit parameters are presented in Table~\ref{tab:phenom_table}, and the fitted spectra are shown in Figure~\ref{fig:phenom_spectra} and the samples of the posterior distributions on the parameters are plotted in Figure~\ref{fig:phenom_contours}. The evolution of the fitted physical parameters is plotted in blue circles in Figure~\ref{fig:evolution}.   The values of the cross-calibration factors are presented in Table~\ref{tab:calib}. The cross-calibration factors obtained can be up to $35\%$ between EPIC and FPM instruments and up to $13\%$ between EPIC cameras. As stated in the calibration technical note\footnote{\url{https://xmmweb.esac.esa.int/docs/documents/CAL-TN-0230-1-3.pdf}} these factors are expected to be high when using the most recent calibration files.\\

The photon index value is usually higher when using \xmm\ data because the data only  extend between $3$ and $10$~keV. The column density of the absorber varies between the two slices of the 2001 observation. It increased from  $0.87\pm0.1$~to $1.4\pm0.2\times10^{24}$~\pcm. A hint of softening in $\Gamma$ can also be seen during this observation. However, the two values are consistent within 1-$\sigma$, increasing from $2.1\pm0.1$ to $2.3\pm0.1$. The observation of 2005 revealed the highest column density $N_\mathrm{H}={2.16}\pm0.16\times10^{24}$~\pcm~and the lowest flux of the reflection component. The two observations of 2012 present the hardest photon index ($\Gamma=1.38\pm{0.05}$ and $\Gamma=1.33\pm0.06$), and the lowest column density ${0.32}\pm{0.04}\times10^{24}$~\pcm. In 2016, the column density decreased continuously from $0.55\pm{0.03}$~to $0.36\pm0.02\times10^{24}$~\pcm~from the start to the end of the observation. In the meantime, the photon index varied as well as the normalisation. Given the low quality of the observation of 2007, only poor constraints on the parameters could be found.

\begin{table*}
\centering

\caption{Best fit values of the phenomenological model across all epochs. The uncertainties are given with a $68\%$ confidence.\\
$^{\rm a}$ Equivalent hydrogen column density modelled with \texttt{TBpcf} in  $10^{22}$~\pcm.\\
$^{\rm b}$ Photon index of the power law.\\
$^{\rm c}$ Normalisation in  $10^{-3}$ ($10^{-5}$) photon~keV$^{-1}$~s$^{-1}$~\pcm~for \texttt{cutoffpl}  (\texttt{xillver}).\\
$^{\rm d}$ Flux of the component in unit of $10^{-12}$~erg~s$^{-1}$~\pcm~either in the 3-10~keV band or in the 10-40~keV band.\\
$^{\rm e}$ Total C-statistic / number of degrees of freedom. }
\renewcommand{\arraystretch}{1.25}
\centering\begin{tabular}{*{10}{l}}\hline\hline
Epoch & ${N_\mathrm{H}}^{~\rm a}$ &$\Gamma^{\rm b}$ & & \texttt{cutoffpl}  & & &\texttt{xillver} &  & cstat/dof$^{~\rm e}$ \\
&  &  & Norm$^{~\rm c}$ & F${_\mathrm{3-10}}^{\rm d}$ &  F${_\mathrm{10-40}}^{~\rm d}$    & Norm$^{\rm c}$  &  F${_\mathrm{3-10}}^{\rm d}$  & F${_\mathrm{10-40}}^{\rm d}$    & \\ \hline
2001-05~{\textcircled{{1}}}&
${86.66}^{+{9.6}}_{-{10.0}}$ &
${2.13}\pm0.1$ &
${6.90}^{+{2.9}}_{-{2.3}}$ &
$10.53_{-2.0}^{+1.8}$ &
$9.94_{-0.9}^{+0.8}$&
${3.56}^{+{0.4}}_{-{0.3}}$ &
$1.95\pm0.1$ &
${7.28}_{-1.6}^{+1.3}$&
$113.84/121$ \\ 
2001-05~{\textcircled{{2}}} 
&${141.81}^{+{15.7}}_{-{14.4}}$ 
&${2.28}\pm0.1$ 
&${14.16}^{+{7.2}}_{-{4.7}}$ 
&$16.90_{-4.5}^{+3.5}$ 
&$13.19_{-2.4}^{+2.0}$
&${4.13}^{+{0.8}}_{-{0.5}}$ 
& $1.91\pm{0.1}$ 
&${5.96}_{-1.2}^{+1.3}$
& $96.40/96$ \\ 
2005-04 &
${216.57}^{+{16.4}}_{-{16.1}}$ &
${1.97}\pm{0.04}$ &
${6.29}^{+{1.5}}_{-{1.2}}$ &
$12.73_{-2.5}^{+2.1}$ &
$14.80_{-2.8}^{+2.3}$ &
${2.67}^{+{0.09}}_{-{0.08}}$ &
$1.61\pm{0.04}$ &
${7.39}\pm0.6$ & 
$251.29/218$ \\ 
2007-04 &
${64.52}^{+{17.8}}_{-{18.2}}$ &
${1.88}^{+{0.5}}_{-{0.6}}$ &
${6.27}^{+{15.2}}_{-{4.4}}$ &
$14.94_{-6.4}^{+3.5}$ &
$19.32_{-12}^{+5.6}$ &
${3.85}^{+{3.7}}_{-{1.2}}$&
$1.52\pm0.4$ &
${8.27}_{-7.2}^{+4.5}$  &
$29.98/29$ \\ 
2012-08 &
${32.36}^{+{4.6}}_{-{4.4}}$ &
${1.38}^{+{0.04}}_{-{0.05}}$ &
${1.22}^{+{0.2}}_{-{0.1}}$  &
$6.91\pm0.6$ &
$17.37\pm1.72$ &
${7.45}^{+{0.7}}_{-{0.6}}$ &
$1.93\pm0.2$ &
${20.20}\pm1.8$ & 
$272.41/227$ \\ 
2012-09 &
${62.82}^{+{11.2}}_{-{9.1}}$ &
${1.33}\pm{0.06}$ &
${0.92}^{+{0.2}}_{-{0.1}}$&
$5.61_{-0.7}^{+0.6}$&
$14.89_{-1.7}^{+1.6}$  &
${7.23}^{+{0.8}}_{-{0.7}}$ &
$1.68\pm0.2$ &
${18.72}_{-1.7}^{+1.8}$ &
$236.79/187$ \\ 
2016-04~{\textcircled{{1}}} &
${55.11}^{+{3.0}}_{-{2.6}}$ &
${1.57}^{+{0.05}}_{-{0.04}}$ &
${2.99}^{+{0.4}}_{-{0.3}}$ &
$12.05_{-0.6}^{+0.5}$ &
$23.50_{-1.1}^{+1.0}$ &
${4.01}\pm0.3$ &
$1.61\pm{0.12}$ &
${12.71}\pm0.9$ & 
$385.72/332$ \\ 
2016-04~{\textcircled{{2}}} &
${50.61}^{+{1.6}}_{-{1.5}}$ &
${1.62}^{+{0.03}}_{-{0.04}}$ &
${6.22}\pm{0.5}$ &
$22.85\pm0.7$ &
${41.81}_{-1.7}^{+1.6}$ &
${3.45}\pm0.3$ &
$1.51\pm{0.2}$ &
${11.20}_{-1.0}^{+1.1}$& 
$419.22/395$ \\ 
2016-04~{\textcircled{{3}}} &
${43.57}^{+{2.0}}_{-{1.8}}$ &
${1.50}\pm{0.04}$ &
${2.83}^{+{0.3}}_{-{0.2}}$ &
$12.78\pm0.5$ 
&${27.24}\pm1.4$ &
${4.61}^{+{0.5}}_{-{0.4}}$ &
$1.61\pm0.1$ &
${13.94}\pm1.2$
& $382.47/356$ \\ 
2016-04~{\textcircled{{4}}} &
${45.84}\pm1.6$ &
${1.51}^{+{0.04}}_{-{0.03}}$ &
${3.77}\pm{0.3}$ &
$16.67\pm0.6$ &
${34.87}_{-1.7}^{+1.6}$
&${3.02}\pm0.4$
&$1.09\pm{0.15}$ 
&${9.23}\pm1.1$
&
$436.85/374$ \\ 
2016-04~{\textcircled{{5}}} 
&${44.40}^{+{1.5}}_{-{1.6}}$ 
&${1.55}\pm{0.03}$ 
&${3.81}\pm0.3$ 
&$15.83_{-0.6}^{+0.5}$ 
&${31.75}_{-1.3}^{+1.2}$
&${3.64}\pm0.3$ 
&$1.41\pm{0.2}$ &
${11.42}\pm1.0$ & 
$470.41/417$ \\ 
2016-04~{\textcircled{6}} &
${42.89}\pm{1.5}$ &
${1.65}\pm{0.03}$ &
${6.38}\pm0.5$ &
$22.41\pm0.9$ &
${39.60}_{-2.0}^{+1.9}$&
${3.68}\pm0.4$ &
$1.68\pm0.2$ &
${11.98}\pm1.4$ & 
$435.84/386$ \\ 
2016-04~{\textcircled{{7}}} &
${35.88}\pm2.2$ &
${1.57}^{+{0.05}}_{-{0.04}}$ &
${5.10}^{+{0.6}}_{-{0.5}}$&
$20.30_{-1.3}^{+1.2}$ &
${39.48}\pm3.0$
&${4.84}\pm0.7$
&$1.95\pm{0.30}$ &
${15.40}\pm2.4$  
&$232.30/227$ \\ \hline\hline
\end{tabular}
\label{tab:phenom_table}
\end{table*}

\subsection{Physical model}
\label{sec:physicalmodel}

Assuming the geometry of the absorber does not change over time, we expect to find a unique geometry fitting to the data for all epochs. We fitted the physical model on all epochs by freezing the geometrical parameters {\tt TORsigma} and {\tt CTKcover}. We performed the fits for eleven out of twenty available geometries because we do not use the geometries which do not include clumps (where ${\tt TORsigma}=0$) or which include large clouds in the thick ring of clouds, i.e. where ${\tt CTKcover}=0.6$.

UltraNest provides an estimate of the marginal likelihood also called evidence $Z$ and samples of the posterior probability distributions. Assuming two models $M_1$ and $M_2$ with respective evidences $Z_1$ and $Z_2$, the Bayes factor $K$ is the ratio of the two marginal likelihoods, i.e. $K={Z_1}/{Z_2}$. \cite{1939thpr.book.....J} gives an empirical interpretation of the values of the Bayes factor, which can be used for model selection.
For instance, if $K<1$, then model $M_1$ is rejected, if $10<K<10^{3/2}$, model $M_1$ is supported with strong evidence over model $M_2$.

Table~\ref{tab:lnZ} shows the natural logarithm of the evidence for the eleven geometries fitted. We sum the logarithm of the evidence over all epochs and use it as a criterion to select the three geometries with the highest $\ln Z$. This rules out the geometries dominated by the inner ring of clouds (with ${\tt TORSigma}=7$) and the partial covering geometry associating clumps and larger thick clouds (${\tt TORSigma}=28$). All the accepted geometries consist of a fully covering clumpy absorber (${\tt TORsigma}=84$). 

\begin{figure*}
    \centering
    \includegraphics[width=\textwidth]{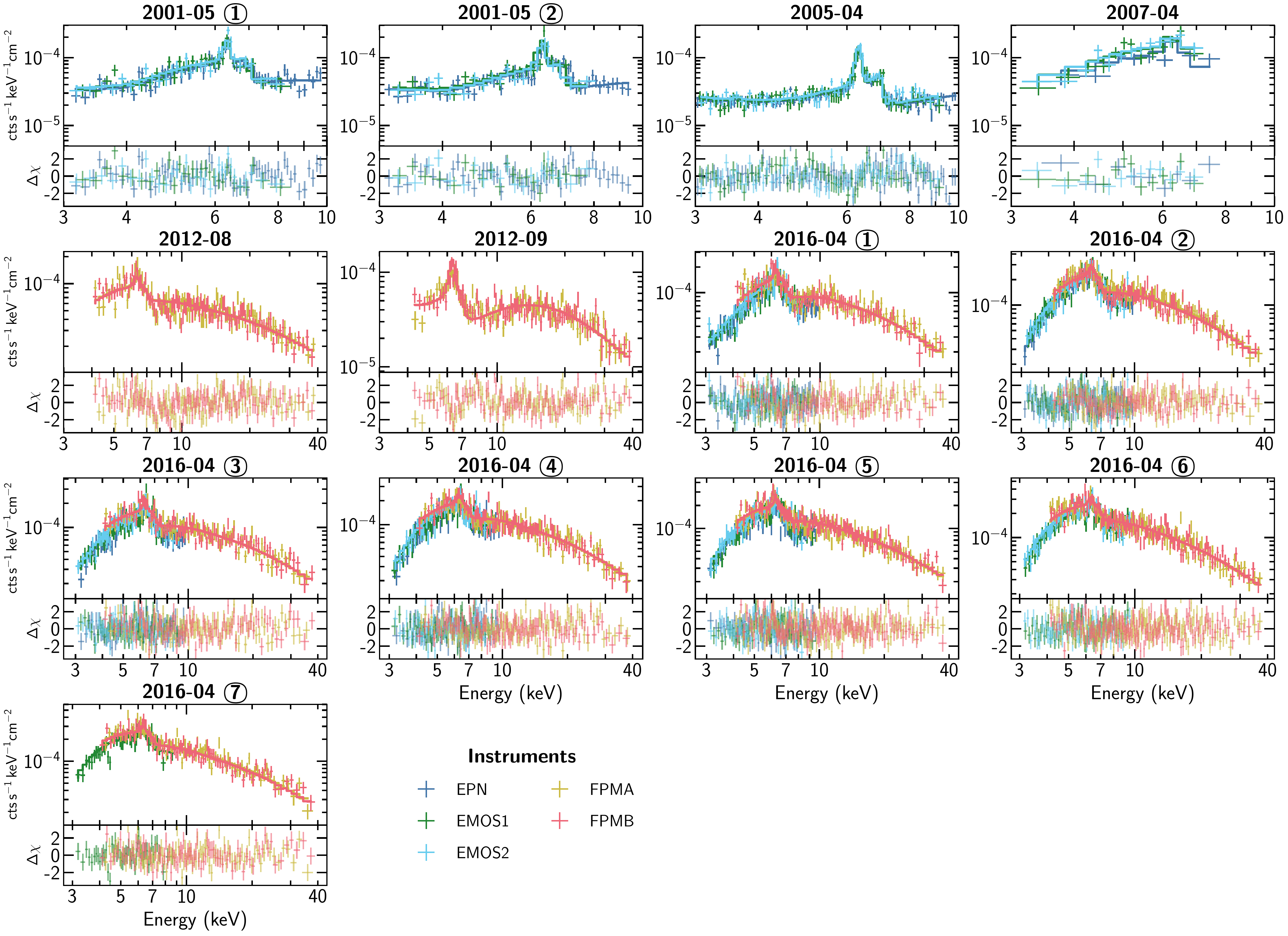}      
    \caption{Best-fit geometry of the fully covering clumpy geometry of the physical model fitted on all epochs with the spectra (upper panels) and residuals (lower panels). The instruments are plotted as follows: EPIC-pn in dark blue, EPIC-MOS1 in green, EPIC-MOS2 in light blue, FPMA in yellow and FPMB in magenta. The circled numbers \textcircled{1}, \textcircled{2}, ...,  \textcircled{7} indicate the slices corresponding to a given observation.}    
    \label{fig:physical_spectra}
\end{figure*}

\begin{table*}
\caption{Best fit values of the fully covering clumpy geometry of the physical model. The uncertainties are given with a $68\%$ confidence.\\
$^{\rm a}$ Equivalent hydrogen column density modelled with \texttt{uxclumpy} in  $10^{22}$~\pcm.\\
$^{\rm b}$ Photon index of the power law.\\
$^{\rm c}$ Normalisation in $10^{-3}$ photon~keV$^{-1}$~s$^{-1}$~\pcm.\\
$^{\rm d}$ Total C-statistic / number of degrees of freedom. }
\renewcommand{\arraystretch}{1.25}
\centering\begin{tabular}{llllll}\hline\hline
Epoch & ${N_\mathrm{H}}^{~\rm a}$ & $\Gamma^{~\rm b}$ & \texttt{uxclumpy} & factor & cstat/dof$^{~\rm d}$ \\
 &  & & Norm$^{~\rm c}$ & $\times 10^{-2}$ & \\\hline
2001-05~{\textcircled{1}} 
&${64.10}\pm{3.7}$ 
&${2.39}\pm0.1$ 
&${32.97}^{+{9.0}}_{-{5.9}}$ 
&${8.79}^{+{0.9}}_{-{1.4}}$ 
& $123.05/123$ \\ 
2001-05~{\textcircled{2}} 
&${82.60}^{+{6.0}}_{-{5.3}}$ 
&${2.25}\pm0.1$
&${29.16}^{+{8.0}}_{-{4.9}}$
&${8.99}^{+{0.8}}_{-{1.3}}$ 
& $94.92/98$ \\ 
2005-04 
&${126.51}^{+{4.5}}_{-{4.6}}$ 
&${2.07}\pm0.1$ 
&${19.79}^{+{7.5}}_{-{4.5}}$
&${6.86}^{+{1.4}}_{-{1.3}}$ 
& $252.08/220$ \\ 
2007-04 
&${37.76}^{+{10.9}}_{-{10.3}}$ 
&${1.88}^{+{0.5}}_{-{0.6}}$ 
&${10.82}^{+{22.1}}_{-{8.0}}$ 
&${6.44}^{+{2.4}}_{-{2.9}}$ 
& $30.11/31$ \\ 
2012-08 
&${20.43}^{+{2.6}}_{-{2.5}}$ 
&${1.33}^{+{0.04}}_{-{0.05}}$ 
&${2.16}^{+{0.4}}_{-{0.3}}$ 
&${6.93}^{+{2.3}}_{-{3.6}}$ 
& $291.17/228$ \\ 
2012-09 
&${34.63}^{+{8.7}}_{-{3.4}}$ 
&${1.30}^{+{0.1}}_{-{0.07}}$ 
&${1.87}^{+{0.9}}_{-{0.3}}$ 
&${7.99}^{+{1.6}}_{-{3.5}}$ 
& $250.03/188$ \\ 
2016-04~{\textcircled{1}} 
&${36.54}^{+{1.4}}_{-{1.5}}$ 
&${1.70}\pm{0.04}$ 
&${7.11}\pm0.8$
&${9.45}^{+{0.4}}_{-{0.8}}$ 
& $379.67/336$ \\ 
2016-04~{\textcircled{2}} 
&${30.26}^{+{1.6}}_{-{1.4}}$ 
&${1.72}^{+{0.03}}_{-{0.04}}$ 
&${11.55}^{+{1.2}}_{-{1.1}}$ 
&${2.73}\pm1.3$
& $429.12/399$ \\ 
2016-04~{\textcircled{3}} 
&${27.80}^{+{1.9}}_{-{1.5}}$
&${1.59}^{+{0.05}}_{-{0.03}}$ 
&${5.69}^{+{0.8}}_{-{0.6}}$
&${8.33}^{+{1.2}}_{-{2.4}} $
& $384.70/360$ \\ 
2016-04~{\textcircled{4}} 
&${27.79}^{+{0.9}}_{-{0.8}}$ 
&${1.62}^{+{0.03}}_{-{0.03}}$ 
&${7.17}^{+{0.6}}_{-{0.5}}$ 
&${2.04}^{+{1.9}}_{-{1.4}}$ 
& $440.63/378$ \\ 
2016-04~{\textcircled{5}} 
&${26.74}^{+{0.9}}_{-{0.8}}$ 
&${1.65}\pm{0.03}$ 
&${7.26}^{+{0.6}}_{-{0.5}}$ 
&${2.05}^{+{2.2}}_{-{1.5}}$ 
& $463.93/421$ \\ 
2016-04~{\textcircled{6}} 
&${26.22}^{+{1.0}}_{-{0.9}}$ 
&${1.75}^{+{0.02}}_{-{0.03}}$ 
&${12.28}\pm{0.9}$ 
&${1.97}^{+{2.5}}_{-{1.4}}$ 
& $433.36/389$ \\ 
2016-04~{\textcircled{7}} 
&${22.07}^{+{1.5}}_{-{1.4}}$ 
&${1.66}\pm{0.05}$ 
&${9.29}^{+{1.3}}_{-{1.0}}$ 
&${4.13}^{+{3.6}}_{-{2.9}}$ 
& $231.85/229$ \\ 
\hline\hline
\end{tabular}
\label{tab:physical_table}
\end{table*}

We compute the Bayes factor for the selected geometries in Table~\ref{tab:bayesphysical} with respect to the best geometry. The Bayes factor is then defined as: $\ln K =\ln Z_{(84,0)}-\ln Z$. Using the Jeffrey scale \citep{1939thpr.book.....J}, we notice that $K>10^2$, hence the fully covering clumpy geometry (${\tt TORSigma}=84$ and ${\tt CTKCover}=0$) is statistically favoured over all geometries and the phenomenological model. 

\begin{table}
\centering
\caption{Values of the natural logarithm of the evidence $Z$ and of the Bayes factor $K$ for the best three geometries and the phenomenological model.\\
The geometries are visualisations of the absorber as seen by the X-ray source. Extracted from the uxclumpy webpage, courtesy of Johannes Buchner.}

\begin{tabular}{llll}\hline\hline
Model & Evidence $\ln Z$ & Bayes factor $\ln K$  \\\hline
(84,0)\,\,\,\,\,\,$\begin{array}{l}\includegraphics[width=0.75cm]{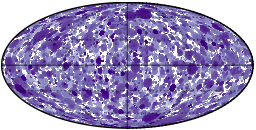}\end{array}$&$-2106$ & - \\
(84,0.2) $\begin{array}{l}\includegraphics[width=0.75cm]{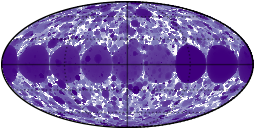}\end{array}$  &$-2111$ & $5$   \\
(84,0.3) $\begin{array}{l}\includegraphics[width=0.75cm]{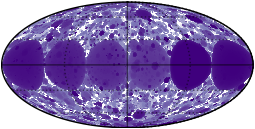}\end{array}$&$-2129$ & $23$  \\
xillver & $-2187$ &$81$  \\
\hline\hline
\end{tabular}
\label{tab:bayesphysical}
\end{table}

Figure~\ref{fig:evolution} shows the best-fit parameters obtained using this geometry in red diamonds and in blue circles the best-fit parameters of the phenomenological model. In Figure~\ref{fig:physical_spectra}, we show this model fitted to all epochs and give the values of the parameters in all epochs in Table~\ref{tab:physical_table}. The samples of the posterior distributions on the parameters for this geometry are plotted in Figure~\ref{fig:physical_contours}.

The values of $\Gamma$ in all epochs are found to be larger than the ones obtained using the phenomenological model, however they are all consistent within 2\,$\sigma$. Values of \NH\ of the physical model are smaller than values of the phenomenological model but they follow the same variations. This is most likely due to the different assumptions of each of the models. The normalisation of \texttt{uxclumpy} follows also the same trend as the one of the \texttt{cutoffpl}.

\section{Discussion}
\label{sec:Discussion}

\subsection{The origin of the variability}

\begin{figure}
    \centering
    \includegraphics[width=\columnwidth]{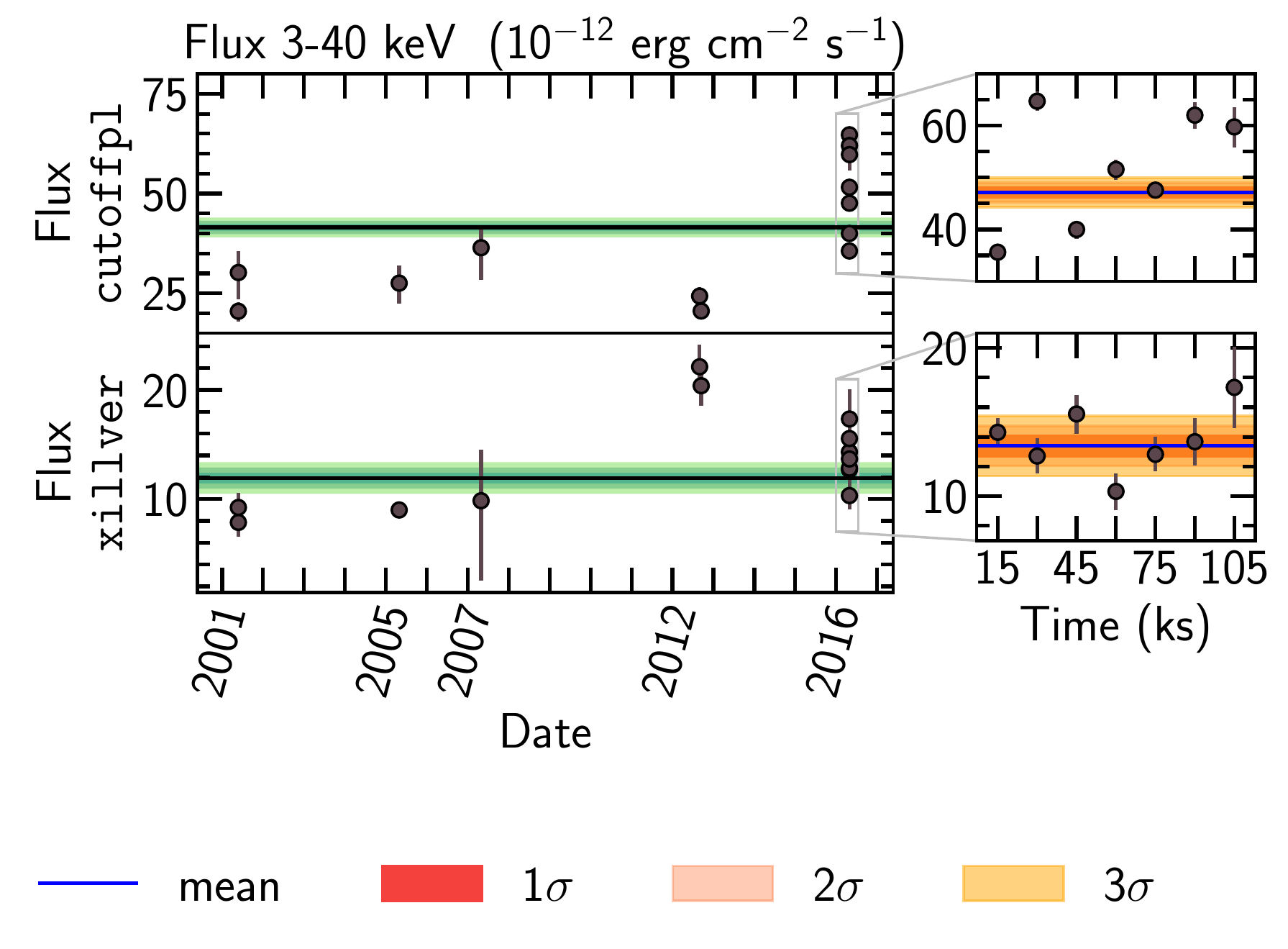}
    \caption{Unabsorbed flux of the power law and reflection components of the phenomenological model in the 3-40~keV band. The fluxes are in units of $10^{-12}$~\flux. The solid lines in each of the panels represent the weighted mean in the given panel and the shaded areas represent the $1\,\sigma$, $2\,\sigma$ and $3\,\sigma$ levels (shades of green in the full panel and red to yellow in the zoomed panel).}
    \label{fig:flux_phenom}
\end{figure}

In Figure~\ref{fig:flux_phenom}, we plot the fluxes in the 3-40 keV energy band for the primary X-ray emission and the reflection component of the phenomenological model. The 10-40 keV fluxes in the \xmm\ observations are obtained by extrapolating the best-fit model to larger energy bands. The fluxes are obtained with {\tt flux} command of XSPEC, on each individual component, i.e \texttt{cutoffpl} and \texttt{xillver}, for all samples of the posterior distribution. The median and the 16${\rm th}$ and 84${\rm th}$ quantiles of the flux distributions are then computed to estimate the $68\%$ confidence levels. We use the phenomenological model as a proxy for distinguishing the contribution of the reflection from one of the power-law.
The separation between these two components cannot be done using our current modelling with \texttt{uxclumpy}. The flux of both components varies over long timescales. However, at short timescale, the reflection component appears to be less variable than the power law (this can be seen in the observation of 2016). This suggests that the response of the reflecting component to the incident primary X-ray emission is not instantaneous. As the observation is $100$~ks long, the reflecting/absorbing material must be located at a distance of  $d>t_\mathrm{obs}\times c \sim 400$~\Rg\ away from the X-ray source. This would put the absorber farther from the black hole, in the BLR.

\begin{figure}
    \centering
    \includegraphics[width=\columnwidth]{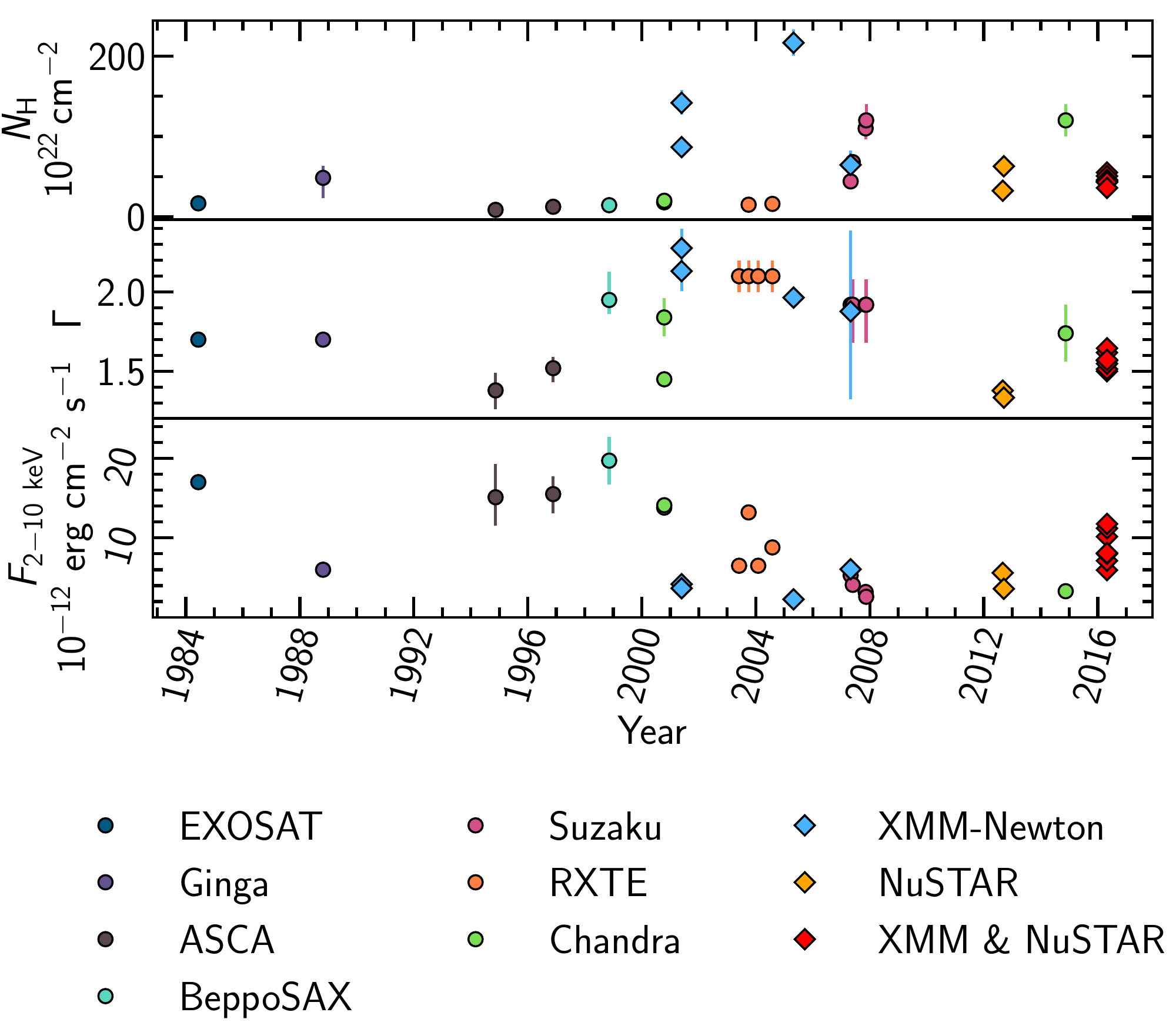}
    \caption{Historical values of the photon index $\Gamma$, the column density \NH\ and the observed flux in the 2-10~keV band for NGC~7582. The values from the literature (available in Table~\ref{tab:histoparams}) are in coloured circles whereas the values obtained with the phenomenological model in this work are in coloured diamonds.}
    \label{fig:historicalvalues}
\end{figure}

Figures~\ref{fig:flux_phenom} and \ref{fig:historicalvalues} demonstrate that the variability on both short- and long-timescales is due to an interplay between intrinsic and absorption-induced variations. NGC~7582 exhibited at least three episodes of nearly Compton-thick state (with $N_{\rm H} > 10^{24}~\rm cm^{-2}$) between 2005 and 2014, which could explain the Changing-look classification of this source. It is worth noting that in 2007, $N_{\rm H}$ may have increased continuously from $4.4_{-0.2}^{+0.3} \times 10^{23}~\rm cm^{-2}$ to $(12 \pm 2)\times 10^{23}~\rm cm^{-2}$ over the course of six months. However, due to the lack of intense monitoring and long observations during this period it is hard to tell whether indeed this rise was continuous or if it also varied on short timescales, similar to the behaviour of the source in 2016. The same reasoning can be applied in 2012, where we observe a decrease of \NH\ between observations separated by fifteen days but with no way to probe the short-term variability.\\

\begin{figure*}
    \centering
    \includegraphics[width=\textwidth]{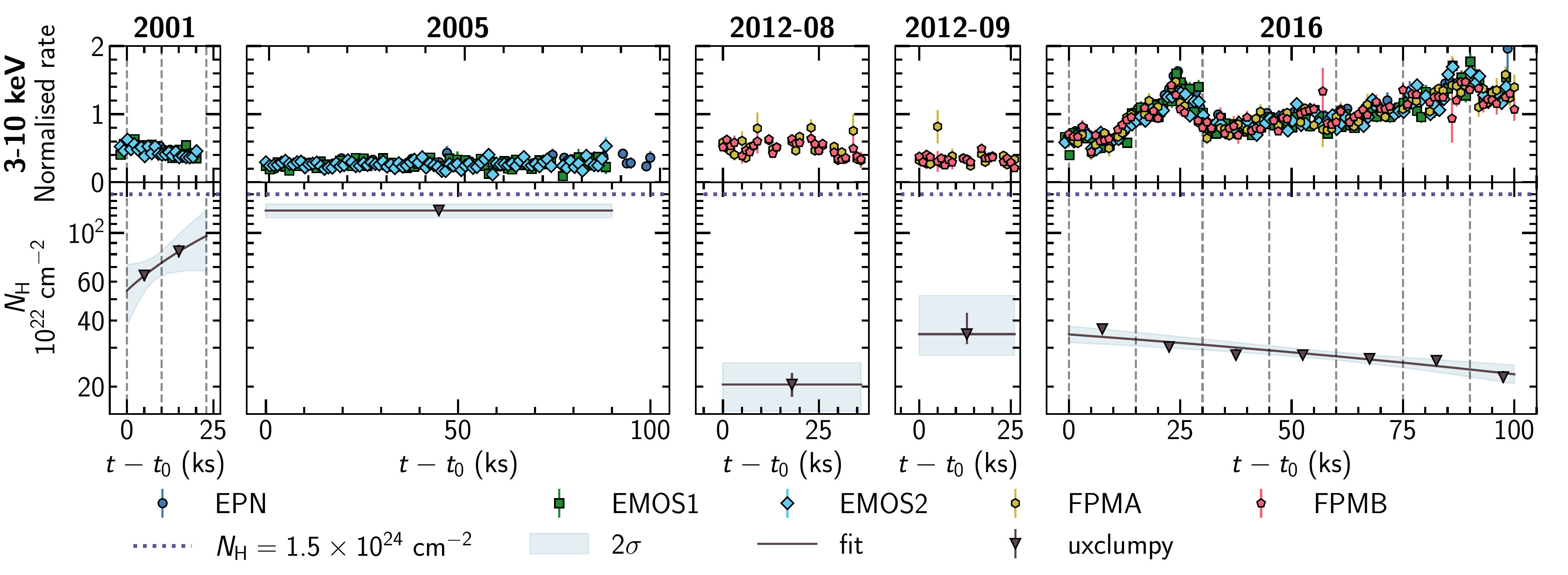}
    \caption{Evolution of the column density of the clumpy absorber (triangles) as a function of time in the observations of 2001, 2005, 2012 and 2016 (bottom panel). The solid line is a linear fit and the shaded area represents the $2\sigma$ level. The Compton-thick column density is delimited by a purple dotted line. The light-curves of \xmm\ EPIC instruments and the \nustar\ FPM instruments in the 3-10~keV band are plotted in the upper-panel, see Figure~\ref{fig:all_lightcurve} for more details.}
    \label{fig:nhrate}
\end{figure*}

In 2001 and 2016, the source exhibited a strong absorption variability. To study this short-term absorption variability, we fitted the evolution of \NH\ with a linear function in these observations using UltraNest. It is worth noting that this model is employed to evaluate the rate at which \NH\ varies without having any physical motivation. The results are presented in Table~\ref{tab:fitnhrate}. As shown in Figure~\ref{fig:nhrate}, we observe a sharp increase in \NH\ in 2001 compared to a slower decrease in 2016. On the contrary, the absorption was constant over 100\,ks during the 2005 observation and over $\sim25$~ks during the observations of 2012. The absence of short-term variability during the 2005 observations suggests that the X-ray emission is absorbed by the same dense absorber during the whole observation. 

The behaviour of the absorber in the different observations can be summarised as follows: a sharp increase in \NH\ (while being Compton thin) in 2001, a constant high \NH\ ($> 10^{24}~\rm cm^{-2}$) absorption over 100\,ks in 2005, an increase of \NH\ between fifteen days in 2012 and a slow decrease in \NH\ over 100\,ks in 2016. These independent episodes could correspond to occultations by various clumps, in agreement with the favoured physical model of a clumpy medium. If the different clumps have similar structures, then these events could agree with a scenario in which the absorbing BLR clouds have a `cometary' structure, with a dense head and a gradually less dense tail. This structure has been proposed to explain the observed eclipsing events in NGC~1365 \citep{2010A&A...517A..47M} and Mrk~766 \citep{2011MNRAS.410.1027R}. If this were true, then the observation of 2001 would correspond to the start of an eclipsing event, the observation of 2005 would correspond to the obscuration by the dense head of the comet, and the observation of 2016 would correspond to the end of an eclipse where the tail of the comet is obscuring the line of sight. However, this scenario cannot be confirmed or ruled out as none of the observations could represent a full eclipse. 

An alternative explanation for the short-term variability is a spherical clump with a density profile peaking in the core and slowly decreasing towards the surface of the clump. The scenario of an eclipse is still valid for a spherical cloud. Thus, longer observations and a closer monitoring may be required to better understand the structure of the absorber in this source.

\begin{table}
    \centering\renewcommand{\arraystretch}{1.25}
    \begin{tabular}{ccc}\hline\hline
    Observation & slope ($10^{22}$~\pcm$\,\mathrm{ks}^{-1}$) & offset ($10^{22}$~\pcm) \\\hline 
        2001-05-25 &  $1.80_{-0.96}^{+0.96}$ & $54.99_{-8.91}^{+9.00}$\\
        2016-04-28 & $-0.12_{-0.02}^{+0.02}$ & $34.58_{-1.51}^{+1.50}$\\\hline\hline
    \end{tabular}
    \caption{Best fit values of the parameters of the linear function fitted onto the evolution of the column density. Errors are given with a $68\%$ confidence level.}
    \label{tab:fitnhrate}
\end{table}

\subsection{Determining the properties of the BLR clouds}
Assuming that the obscuring clouds are located in the {BLR}, following a Keplerian orbit, we derive below some of the properties of these clouds.

Velocity: \cite{1999ApJ...519L.123A} observed NGC~7582 in a Seyfert-1 state with broad optical lines on the 11 July 1998. The estimated FWHM of the H$\alpha$ and H$\beta$ lines is FWHM$=12400$~\kmps; over the timescale of a few months the FWHM decreased to reach $6660$~\kmps\, at the end of October 1998 \citep{1999ApJ...519L.123A}. \cite{2018MNRAS.473.5334R} reported a FWHM of $2382\pm44$~\kmps, in an observation of 2004. With a 2014 \textit{Chandra} observation, \cite{2017AA...600A.135B} estimated the FWHM of the Fe K$\alpha$ line to be FWHM$=1500_{-800}^{+900}$~\kmps, thus $v=637$~\kmps. Because no measurement of the FWHM of H$\alpha$ was made close to our observations, it is impossible to predict the state of the source in the optical at the time of  our archival observations. We tried to estimate the FWHM of the Fe K$\alpha$ in our \xmm\ observations but the uncertainties were too high to draw any conclusions. We use the most recent measurement by \cite{2017AA...600A.135B} to provide a lower limit on the velocity of the clouds: $v\geq637$~\kmps.\\

\rm Size of the clouds: As we do not observe complete eclipsing events, we can only set lower limits on the size of the clouds. The typical duration of an eclipse is at least twice the exposure time $t_\mathrm{exp}=100$~ks. Assuming a spherical cloud of diameter $D_c$ moving at a velocity $v_c$, the duration of an eclipse is $t_\mathrm{ec}=D_c/v_c$.  The diameter of the cloud is then $D_c\geq 2v_c\times t_\mathrm{exp}$. Using the value of velocity obtained above, $D_c\gtrsim1.5$~\Rg~($1.2\times 10^{13}$~cm).\\

Location of the BLR: Assuming a Keplerian circular orbit for the clumps in the {BLR}. The radius of the orbit in the BLR is given by $R=\frac{GM}{v^2}=c^2/v^2$~\Rg, this  gives an upper limit on the location of the clouds in the BLR, $R_{\rm BLR}/R_{g} \lesssim 22000$, $R_{\rm BLR}\sim 0.6$~pc.

\subsection{Comparison with previous works}

NGC~7582 has been observed by many X-ray observatories. In Figure~\ref{fig:historicalvalues}, we plot the values of the column density, the photon index, and the observed flux obtained from previous studies (in circles) and the ones obtained in this work (in diamonds). The previous works mostly used phenomenological models, thus we show the parameters obtained with our best-fit phenomenological model, for consistency. The long-term variability is visible since early observations. As seen in this figure, \textit{BeppoSAX} observed NGC~7582 in an unobscured state, as opposed to the very absorbed state of 2005 seen by \textit{XMM-Newton}. We provide a non-exhaustive list of values from the literature in Table~\ref{tab:histoparams}, where we did not include results from the five observations with the \textit{Einstein} observatory as it was not possible to get access to all the fitted parameters. Below, we present a comparison between the values reported in the literature to the ones obtained in this work.

{\rm Observation 2001:} For this observation, the values of the photon index obtained in this work are consistent with the values obtained by \cite{2005ApJ...625L..31D} and \cite{2007AA...466..855P}. The value of \NH\ obtained in the two works are in the same order of magnitude as the value found in slice~\textcircled{1}. It is worth noting that \cite{2007AA...466..855P} used a two-absorber model. Thus, we compare only with the highest column density, as the fitting range we adopt ($E \geq 3\,\rm keV$) is not sensitive to the low column density absorber.

{\rm Observation 2005:} In this observation, the column density and photon index obtained by \cite{2007AA...466..855P} are also consistent with the ones obtained in this work. \cite{2020ApJ...897...66L} fitted a double-absorber model on this observation and the total column density as well as the photon index are consistent with our work.

{\rm Observation 2007:} \cite{2009ApJ...695..781B} fitted simultaneously the \textit{Suzaku} observations with the \xmm\ observation of 2007. Given the poor quality of our data the values are poorly constrained and {de facto} consistent with values from the literature \citep[e.g., ][]{2020ApJ...897...66L}.

{\rm Observation 2012:} For the \textit{NuSTAR} observations of 2012, the values of \NH\ are consistent with the findings of \cite{2015ApJ...815...55R}. However, the values of $\Gamma$ are smaller than the ones found previously ($1.33$ against $1.78$). It worth mentioning that they fitted simultaneously the first observation with a \textit{Swift} observation and tied several parameters between the two observations of 2012.

{\rm Observation 2016:} \cite{2022ApJS..260...30T} applied a physical clumpy model  \citep[XCLUMPY;][]{2019ApJ...877...95T} to the time-averaged spectrum of the observation of 2016. \cite{2018ApJ...854...42B} applied a torus model to the \nustar\ observation only. The values of $\Gamma$ and \NH\ we obtain on average are consistent with the ones obtained by both \cite{2022ApJS..260...30T} and \cite{2018ApJ...854...42B}. These works found a covering factor of $80-90\%$ which is roughly consistent with our full covering geometry. It is worth noting that these papers did not study the short-term variability seen in this observation.

\subsection{Prospects with future X-ray missions}

\begin{figure*}
     \centering
     \begin{subfigure}[c]{0.58\textwidth}
         \centering
         \includegraphics[width=\textwidth]{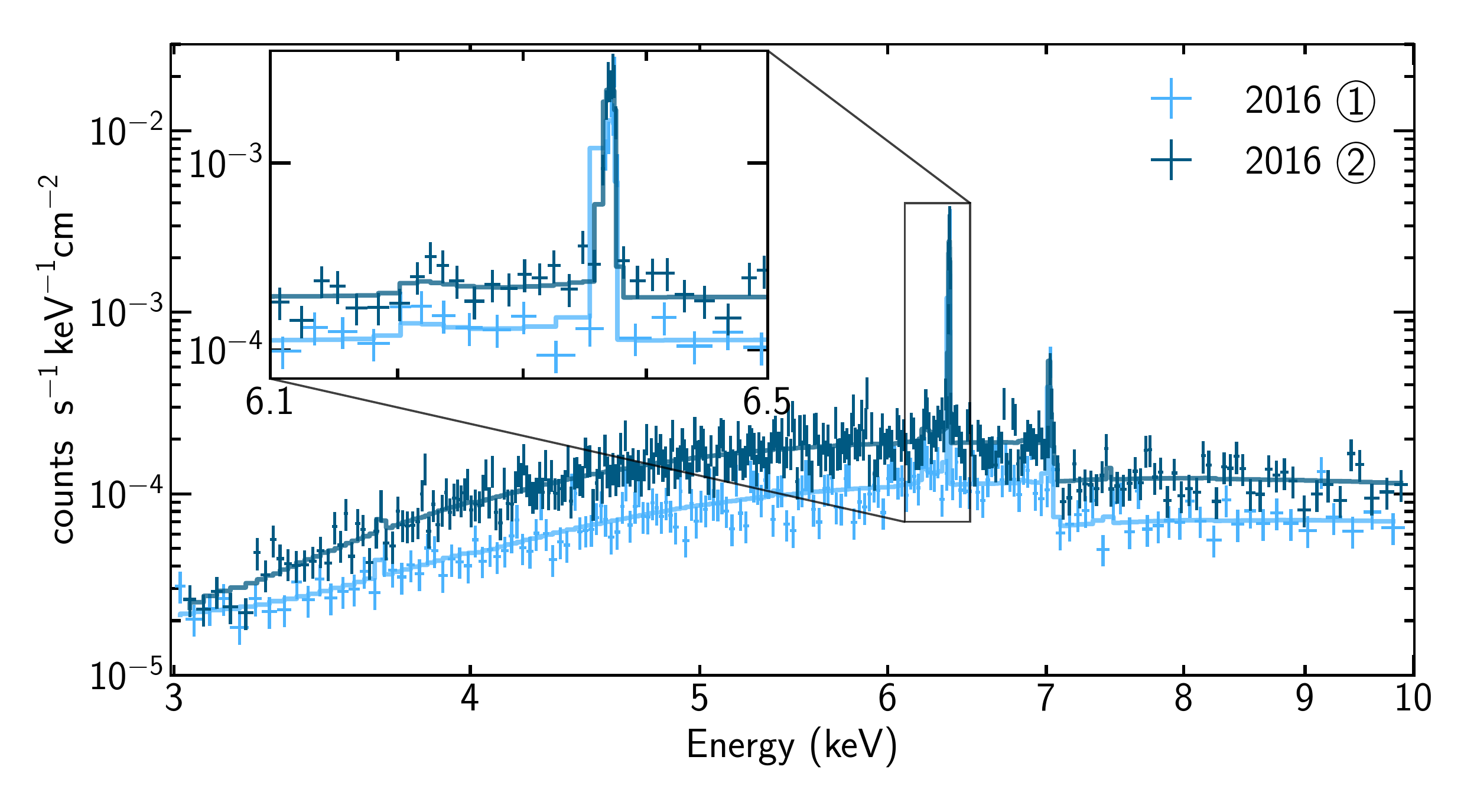}
         \caption{\textit{Athena}/X-IFU}
         \label{fig:xifu}
     \end{subfigure}
     \hfill
     \begin{subfigure}[c]{0.4\textwidth}
         \centering
         \includegraphics[width=\textwidth]{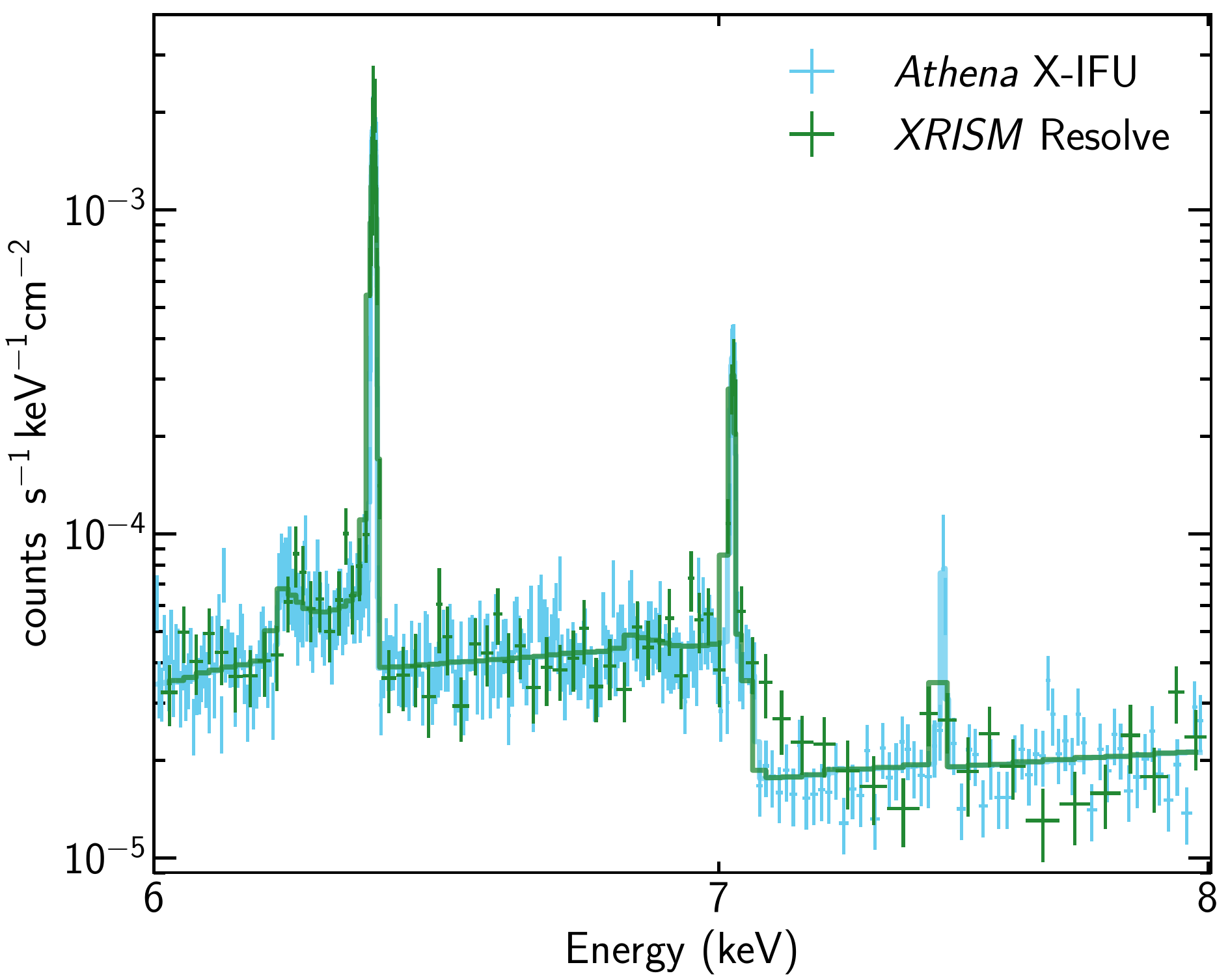}
         \caption{\textit{Athena}/X-IFU and \textit{XRISM}/Resolve.}
         \label{fig:xrism}
     \end{subfigure}
        \caption{Simulations of spectra with microcalimeter detectors assuming the best-fit \texttt{uxclumpy} model. On the left, simulations of the first two slices of 2016, using \textit{Athena}/X-IFU with an exposure time of 5~ks. The simulated spectrum of the first slice is in light blue and the second slice is in dark blue. On the right, simulations of the observation of 2005, using \textit{Athena}/X-IFU (in cyan) \textit{XRISM}/Resolve (in green) with an exposure time of 90~ks. The spectra are shown in the 6-8~keV range. The inset shows a zoom in the 6.1-6.5~kev range where the Fe K$\alpha$ and the Compton shoulder are resolved.}
        \label{fig:simulationsxifuxrism}
\end{figure*}

\begin{figure}
    \centering
    \includegraphics[width=\columnwidth]{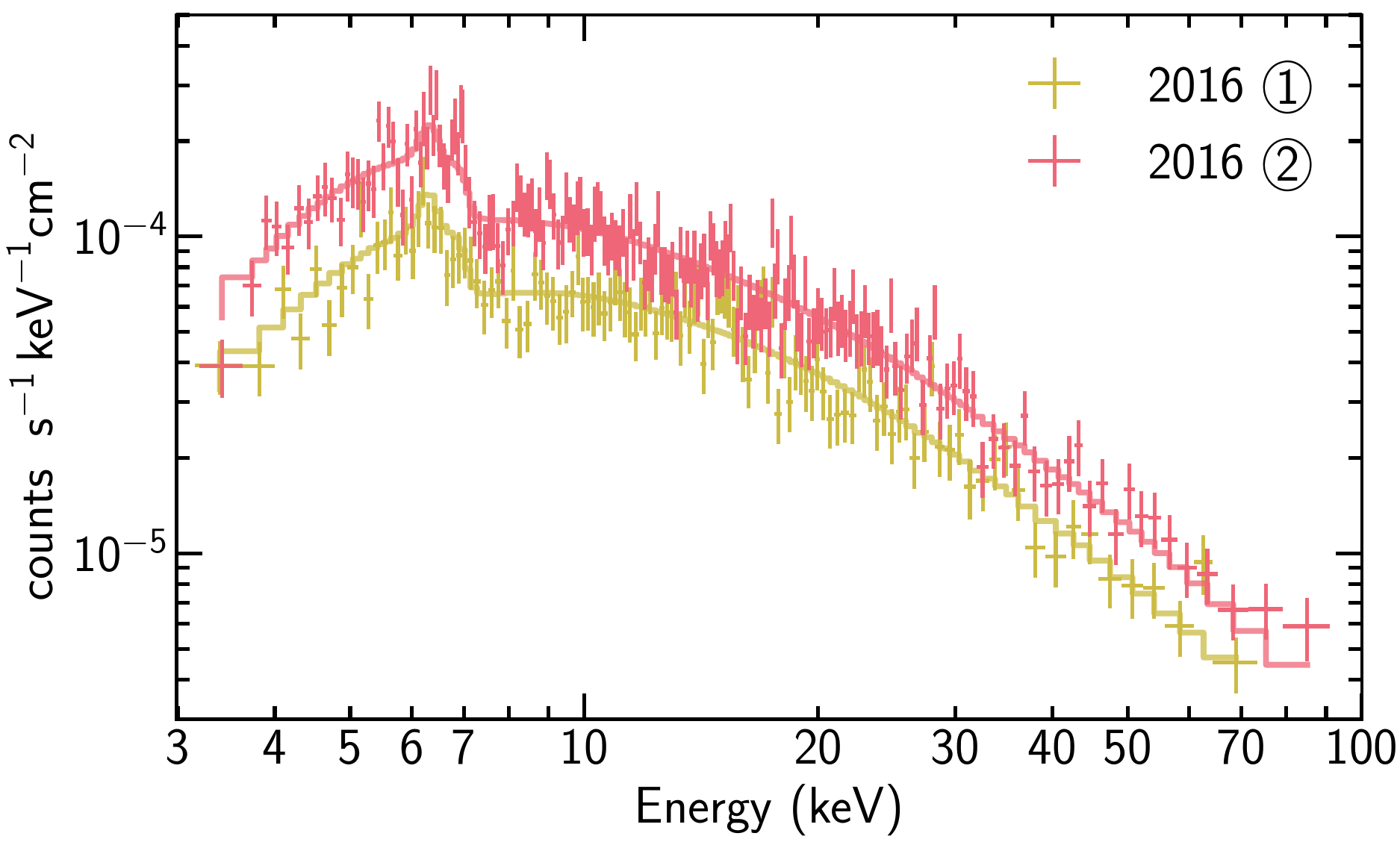}
    \caption{Simulated \textit{HEX-P}/HET spectra of the first two slices of 2016 with an exposure time of 5~ks assuming the best-fit \texttt{uxclumpy} model.}
    \label{fig:hexp}
\end{figure}

Future X-ray missions will bring the opportunity to improve the current work and better constrain the physical parameters. We simulate data for several  instruments assuming the best-fit fully covering model with \texttt{uxclumpy} and we rebin all spectra with at least 25~counts per bin. 
In Figure~\ref{fig:xifu}, we show two simulated \textit{Athena}/X-IFU \citep{Barret2022} spectra for the observation of 2016 with an exposure time of 5~ks each. Thanks to the large effective area and the high resolution of \textit{Athena}/X-IFU\footnote{\url{http://x-ifu.irap.omp.eu/}}, it will be possible to probe the spectral variability on shorter timescales with better constraints. In Figure~\ref{fig:xrism}, we show simulated spectra for the observation of 2005 with an exposure time of 90\,ks with \textit{Athena}/X-IFU and \textit{XRISM}/Resolve\footnote{\url{https://heasarc.gsfc.nasa.gov/docs/xrism/}} \citep{2020SPIE11444E..22T}. The inset of Figure~\ref{fig:xifu} and Figure~\ref{fig:xrism} show a view in the 6.1-6.5 keV range. The Fe K$\alpha$ doublet at 6.404 keV and 6.391 keV cannot be resolved, nevertheless the Compton shoulder can be observed with the 90-ks simulations.  \textit{Athena} and \textit{XRISM} will enable high-resolution spectroscopy of the low-luminosity state of the AGN as observed in the observation of 2005. Finally, with the High-Energy Telescope (HET) of the concept mission \textit{HEX-P}\footnote{\url{https://hexp.org}} \citep{2018SPIE10699E..6MM}, observations above 10~keV could constrain both the photon index and the high-energy cutoff. In Figure~\ref{fig:hexp}, we show two simulated HET spectra for the observation of 2016 with an exposure time of 5~ks each. Thus, each of these three instruments will enable a better understanding of both the accretion in AGN but also the geometry of the absorbing material surrounding the X-ray source.

\section{Conclusions}
\label{sec:Conclusion}
To model the spectra we have used two different models, a physically-motivated model accounting for the geometry of the absorber and a phenomenological model. The main results can be summarised as follows:
\begin{itemize}
    \item The absorption is best described by a fully covering clumpy absorber. This geometry is statistically favoured using Bayes factors.
    \item Long-term and short-term variability are due to intrinsic changes in the X-ray luminosity of the source, and changes in the absorption in the line of sight. The latter can explain the changing-look classification of this source, more specifically changing-obscuration, as the source switches between Compton-thick and Compton-thin due to changes in the obscuring column along the line of sight, respectively.
    \item The absorber is located at a distance not larger than $0.6$~pc.
    \item Assuming a single cloud in the line of sight, we estimate the size of the cloud to be larger than $10^{13}$~cm, for a velocity of $\sim700$~\kmps.
    \item Given the variety of behaviour of the absorption at different timescales, and the fact that the observations are not long enough to constrain a full eclipsing event, we cannot make strong arguments on the shape of the clouds. However, our results strongly suggest the presence of a variety of cloud densities, or even a gradient in the density within a single cloud (as seen in 2016), favouring comet-shaped clouds or at least a non-uniform cloud.
\end{itemize}


\section*{Acknowledgements}

We thank the anonymous referee for their comments and suggestions. ML acknowledges useful discussions with Johannes Buchner on the UXCLUMPY model. DB/ESK/ML acknowledge financial support from the Centre National d’Etudes Spatiales (CNES). ML acknowledges financial support through the UK Science and Technology Facilities Council (STFC). AZ is supported by NASA under award number 80GSFC21M0002.

This research has made use of data and/or software provided by the High Energy Astrophysics Science Archive Research Center (HEASARC), which is a service of the Astrophysics Science Division at NASA/GSFC and the High Energy Astrophysics Division of the Smithsonian Astrophysical Observatory. This research made use of Astropy, a community-developed core Python package for Astronomy \citep{2018AJ....156..123A, 2013A&A...558A..33A}, NumPy \citep{harris2020array} and matplotlib, a Python library for publication quality graphics \citep{Hunter:2007}. This work has made use of GetDist \citep{2019arXiv191013970L}, a Python library for analysing Monte Carlo samples. This research made use of XSPEC \citep{1996ASPC..101...17A}.
We have made use of data from the NuSTAR mission, a project led by the California
Institute of Technology, managed by the Jet Propulsion
Laboratory, and funded by the National Aeronautics and Space
Administration. NuSTAR data was reduced with the NuSTAR Data Analysis Software (NuSTARDAS) jointly developed by the ASI Science Data Center (ASDC, Italy) and the California Institute of Technology (USA). We have also made use of data based on observations obtained with XMM-Newton, an ESA science mission with instruments and contributions directly funded by ESA Member States and NASA.


\section*{Data Availability}

The data used in this paper are publicly available to access and download from the High Energy Astrophysics Science Archive Research Center for NuSTAR and from the XMM-Newton Science Archive. Final data products from this study can be provided on reasonable request to the corresponding author.
 



\bibliographystyle{mnras}
\bibliography{references} 




\appendix

\section{Tables and Figures}
\label{sec:crosscalib}
\begin{table}
 \centering\caption{Calibration factors across all epochs for all instruments. The uncertainties are given with a $68\%$ confidence.}
\renewcommand{\arraystretch}{1.25}
\begin{tabular}{lccccc}\hline\hline
Epoch & & &Instruments & &   \\  
         & EPN & EMOS1 & EMOS2  & FPMA  & FPMB  \\\hline
2001-05~{\textcircled{1} }                             & 1   & $1.08_{-0.05}^{+0.04}$ & $1.08_{-0.04}^{+0.05}$ &    -   &   -    \\
2001-05~{\textcircled{{2}}}                                          & 1   & $1.08_{-0.05}^{+0.05}$ & $1.02_{-0.05}^{+0.05}$ &    -   &    -   \\
2005-04                                      & 1  & $1.04_{-0.02}^{+0.02}$  & $1.08_{-0.02}^{+0.02}$ &  -     &    -   \\
2007-04                                    & 1   & $1.19_{-0.09}^{+0.09}$  & $1.25_{-0.08}^{+0.06}$    &    -   &    -   \\
2012-08                 &   -  &    -   &  -      & 1     & $0.98_{-0.02}^{+0.02}$  \\
2012-09                         &  -   &    -   &  -      & 1     & $1.00_{-0.03}^{+0.03}$  \\
2016-04~{\textcircled{{1}}}      & 1   & $1.04_{-0.03}^{+0.03}$  & $1.10_{-0.03}^{+0.03}$  & $1.31_{-0.03}^{+0.03}$  & $1.28_{-0.04}^{+0.03}$  \\
2016-04~{\textcircled{{2}}}      & 1   & $1.09_{-0.03}^{+0.03}$  & $1.06_{-0.02}^{+0.03}$  & $1.26_{-0.03}^{+0.03}$  & $1.20_{-0.03}^{+0.03}$ \\
2016-04~{\textcircled{{3}}}      & 1   & $1.02_{-0.03}^{+0.03}$  & $1.09_{-0.03}^{+0.04}$  & $1.27_{-0.04}^{+0.04}$  & $1.27_{-0.04}^{+0.04}$ \\
2016-04~{\textcircled{{4}}}      & 1   & $1.04_{-0.03}^{+0.03}$  & $1.11_{-0.03}^{+0.03}$  & $1.26_{-0.04}^{+0.04}$  & $1.28_{-0.04}^{+0.04}$ \\
2016-04~{\textcircled{{5}}}      & 1   & $1.08_{-0.03}^{+0.03}$  & $1.13_{-0.03}^{+0.03}$  & $1.29_{-0.03}^{+0.03}$  & $1.34_{-0.02}^{+0.01}$ \\
2016-04~{\textcircled{{6}}}      & -    & 1     &  $1.06_{-0.03}^{+0.03}$ &  $1.24_{-0.03}^{+0.04}$ &  $1.25_{-0.04}^{+0.04}$ \\
2016-04~{\textcircled{{7}}}     & -    & 1     &     -   &  $1.25_{-0.05}^{+0.05}$  &  $1.25_{-0.04}^{+0.05}$ \\\hline\hline
\end{tabular}
\label{tab:calib}
\end{table}


\begin{figure*}
    \centering
    \begin{center} 
    \includegraphics[width=\textwidth]{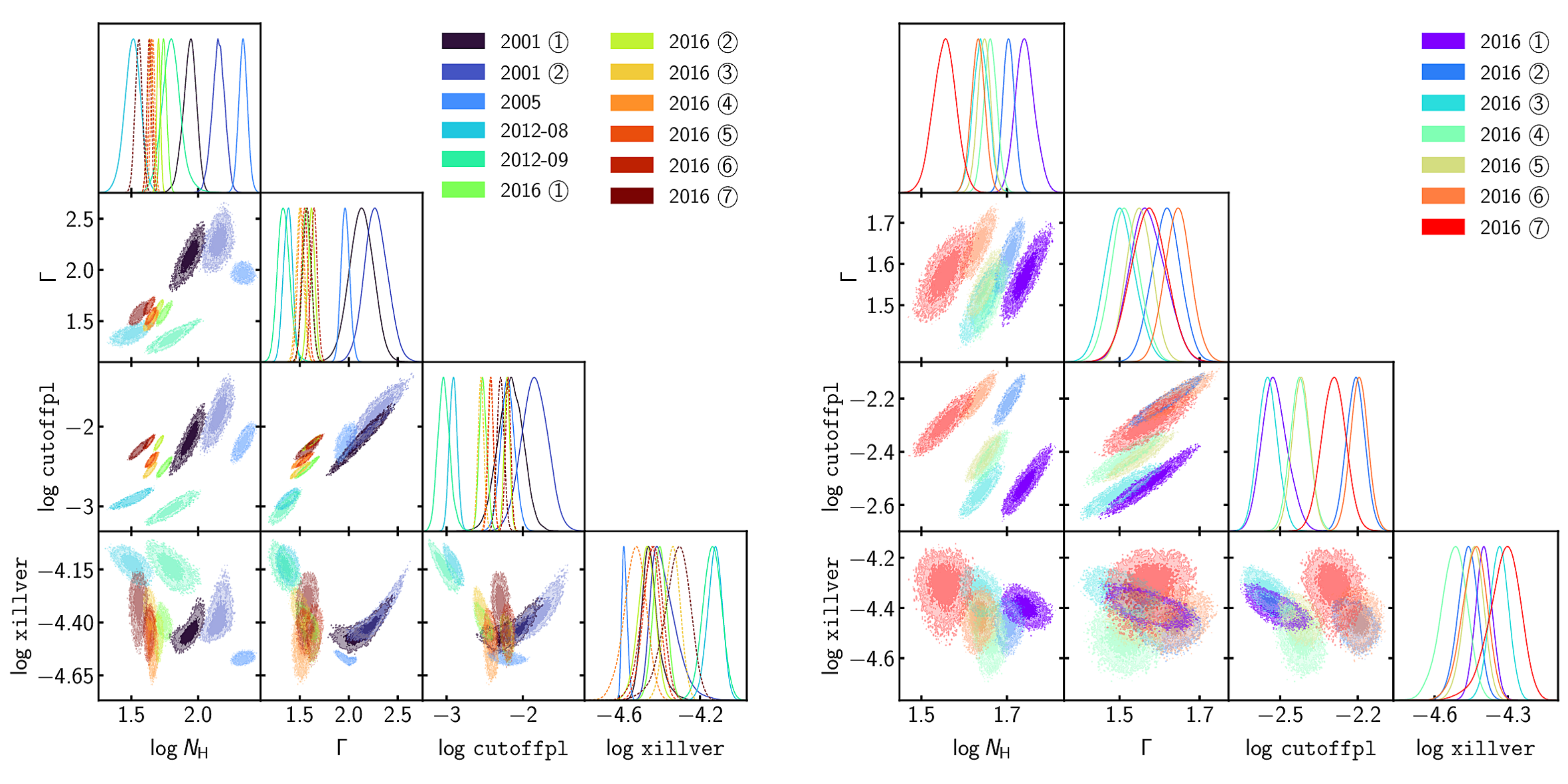}
 \end{center}
    \caption{Contours of the samples of the posterior distributions of the parameters of the phenomenological model. For all epochs (left) and all slices of the observation of 2016 (right).}
    \label{fig:phenom_contours}
\end{figure*}

\begin{table*}
    \centering
    \caption{Values of $\ln Z$ for eleven geometries of the physical model and the phenomenological model over all epochs. The gradient of colour shows for each row maximal values (dark blue) to minimal (yellow). The last row gives the total $\ln Z$ for all epochs. The geometries are visualisations of the absorber as seen by the X-ray source. Extracted from the uxclumpy webpage, courtesy of Johannes Buchner.}
\label{tab:lnZ}
\begin{tabular}{lrrrrrrrrrrrr}\hline\hline
 \multicolumn{1}{l}{\multirow{3}{*}{\textbf{Epoch}}} &
 \multicolumn{1}{l}{\multirow{3}{*}{\textbf{xillver}}}&\multicolumn{11}{c}{\textbf{Geometries \texttt{\textbf{uxclumpy}} ~~ ({\tt TORsigma, CTKcover})}}      \\  
\multicolumn{1}{l}{}  &    \multicolumn{1}{l}{}  &   \multicolumn{1}{c}{(7,0)} &\multicolumn{1}{c}{(7,0.3)} &\multicolumn{1}{c}{(7,0.5)} &\multicolumn{1}{c}{(28,0)} &\multicolumn{1}{c}{(28,0.2)} &\multicolumn{1}{c}{(28,0.3)} &\multicolumn{1}{c}{(28,0.5)} &\multicolumn{1}{c}{(84,0)} &\multicolumn{1}{c}{(84,0.2)} &\multicolumn{1}{c}{(84,0.3)} &\multicolumn{1}{c}{(84,0.5)} \\
\multicolumn{1}{l}{}  &    \multicolumn{1}{l}{}  &   \includegraphics[width=1cm]{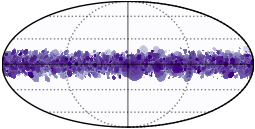}  & \includegraphics[width=1cm]{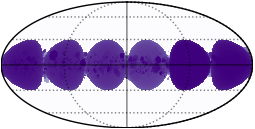}  &\includegraphics[width=1cm]{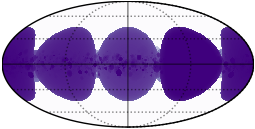}  & \includegraphics[width=1cm]{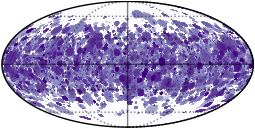}  &\includegraphics[width=1cm]{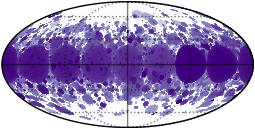}  & \includegraphics[width=1cm]{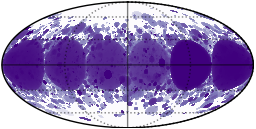}  & \includegraphics[width=1cm]{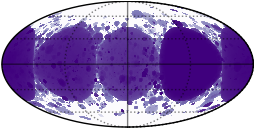}  & \includegraphics[width=1cm]{figures/appendix/geometries/84-0.png} & \includegraphics[width=1cm]{figures/appendix/geometries/84-0.2.png}  &\includegraphics[width=1cm]{figures/appendix/geometries/84-0.3.png}  & \includegraphics[width=1cm]{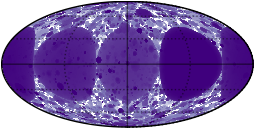} 
\\
\hline
2001~{\textcircled{{1}}} & {\cellcolor[HTML]{002553}} \color[HTML]{F1F1F1} {\cellcolor[HTML]{002553}} \color[HTML]{F1F1F1} -77 & {\cellcolor[HTML]{8D8878}} \color[HTML]{F1F1F1} {\cellcolor[HTML]{8D8878}} \color[HTML]{F1F1F1} -97 & {\cellcolor[HTML]{FEE838}} \color[HTML]{000000} {\cellcolor[HTML]{FEE838}} \color[HTML]{000000} -113 & {\cellcolor[HTML]{DAC65B}} \color[HTML]{000000} {\cellcolor[HTML]{DAC65B}} \color[HTML]{000000} -108 & {\cellcolor[HTML]{31436D}} \color[HTML]{F1F1F1} {\cellcolor[HTML]{31436D}} \color[HTML]{F1F1F1} -83 & {\cellcolor[HTML]{51586D}} \color[HTML]{F1F1F1} {\cellcolor[HTML]{51586D}} \color[HTML]{F1F1F1} -88 & {\cellcolor[HTML]{717274}} \color[HTML]{F1F1F1} {\cellcolor[HTML]{717274}} \color[HTML]{F1F1F1} -93 & {\cellcolor[HTML]{C7B767}} \color[HTML]{000000} {\cellcolor[HTML]{C7B767}} \color[HTML]{000000} -106 & {\cellcolor[HTML]{00224E}} \color[HTML]{F1F1F1} {\cellcolor[HTML]{00224E}} \color[HTML]{F1F1F1} -76 & {\cellcolor[HTML]{053371}} \color[HTML]{F1F1F1} {\cellcolor[HTML]{053371}} \color[HTML]{F1F1F1} -80 & {\cellcolor[HTML]{414D6C}} \color[HTML]{F1F1F1} {\cellcolor[HTML]{414D6C}} \color[HTML]{F1F1F1} -85 & {\cellcolor[HTML]{999277}} \color[HTML]{F1F1F1} {\cellcolor[HTML]{999277}} \color[HTML]{F1F1F1} -99 \\
2001~{\textcircled{{2}}} & {\cellcolor[HTML]{50576C}} \color[HTML]{F1F1F1} {\cellcolor[HTML]{50576C}} \color[HTML]{F1F1F1} -68 & {\cellcolor[HTML]{A59C74}} \color[HTML]{F1F1F1} {\cellcolor[HTML]{A59C74}} \color[HTML]{F1F1F1} -74 & {\cellcolor[HTML]{FEE838}} \color[HTML]{000000} {\cellcolor[HTML]{FEE838}} \color[HTML]{000000} -80 & {\cellcolor[HTML]{F0D846}} \color[HTML]{000000} {\cellcolor[HTML]{F0D846}} \color[HTML]{000000} -79 & {\cellcolor[HTML]{002656}} \color[HTML]{F1F1F1} {\cellcolor[HTML]{002656}} \color[HTML]{F1F1F1} -62 & {\cellcolor[HTML]{2F426D}} \color[HTML]{F1F1F1} {\cellcolor[HTML]{2F426D}} \color[HTML]{F1F1F1} -65 & {\cellcolor[HTML]{595E6E}} \color[HTML]{F1F1F1} {\cellcolor[HTML]{595E6E}} \color[HTML]{F1F1F1} -68 & {\cellcolor[HTML]{DDC858}} \color[HTML]{000000} {\cellcolor[HTML]{DDC858}} \color[HTML]{000000} -78 & {\cellcolor[HTML]{00224E}} \color[HTML]{F1F1F1} {\cellcolor[HTML]{00224E}} \color[HTML]{F1F1F1} -62 & {\cellcolor[HTML]{163770}} \color[HTML]{F1F1F1} {\cellcolor[HTML]{163770}} \color[HTML]{F1F1F1} -64 & {\cellcolor[HTML]{47516C}} \color[HTML]{F1F1F1} {\cellcolor[HTML]{47516C}} \color[HTML]{F1F1F1} -67 & {\cellcolor[HTML]{918B78}} \color[HTML]{F1F1F1} {\cellcolor[HTML]{918B78}} \color[HTML]{F1F1F1} -73 \\
2005 & {\cellcolor[HTML]{0F3570}} \color[HTML]{F1F1F1} {\cellcolor[HTML]{0F3570}} \color[HTML]{F1F1F1} -149 & {\cellcolor[HTML]{8E8978}} \color[HTML]{F1F1F1} {\cellcolor[HTML]{8E8978}} \color[HTML]{F1F1F1} -177 & {\cellcolor[HTML]{E0CB56}} \color[HTML]{000000} {\cellcolor[HTML]{E0CB56}} \color[HTML]{000000} -196 & {\cellcolor[HTML]{FEE838}} \color[HTML]{000000} {\cellcolor[HTML]{FEE838}} \color[HTML]{000000} -204 & {\cellcolor[HTML]{46516C}} \color[HTML]{F1F1F1} {\cellcolor[HTML]{46516C}} \color[HTML]{F1F1F1} -159 & {\cellcolor[HTML]{777776}} \color[HTML]{F1F1F1} {\cellcolor[HTML]{777776}} \color[HTML]{F1F1F1} -171 & {\cellcolor[HTML]{898578}} \color[HTML]{F1F1F1} {\cellcolor[HTML]{898578}} \color[HTML]{F1F1F1} -176 & {\cellcolor[HTML]{F8DF3C}} \color[HTML]{000000} {\cellcolor[HTML]{F8DF3C}} \color[HTML]{000000} -201 & {\cellcolor[HTML]{00224E}} \color[HTML]{F1F1F1} {\cellcolor[HTML]{00224E}} \color[HTML]{F1F1F1} -142 & {\cellcolor[HTML]{002A61}} \color[HTML]{F1F1F1} {\cellcolor[HTML]{002A61}} \color[HTML]{F1F1F1} -145 & {\cellcolor[HTML]{123570}} \color[HTML]{F1F1F1} {\cellcolor[HTML]{123570}} \color[HTML]{F1F1F1} -149 & {\cellcolor[HTML]{4F576C}} \color[HTML]{F1F1F1} {\cellcolor[HTML]{4F576C}} \color[HTML]{F1F1F1} -161 \\
2007 & {\cellcolor[HTML]{FEE838}} \color[HTML]{000000} {\cellcolor[HTML]{FEE838}} \color[HTML]{000000} -30 & {\cellcolor[HTML]{35456C}} \color[HTML]{F1F1F1} {\cellcolor[HTML]{35456C}} \color[HTML]{F1F1F1} -27 & {\cellcolor[HTML]{4E566C}} \color[HTML]{F1F1F1} {\cellcolor[HTML]{4E566C}} \color[HTML]{F1F1F1} -27 & {\cellcolor[HTML]{33446D}} \color[HTML]{F1F1F1} {\cellcolor[HTML]{33446D}} \color[HTML]{F1F1F1} -27 & {\cellcolor[HTML]{002553}} \color[HTML]{F1F1F1} {\cellcolor[HTML]{002553}} \color[HTML]{F1F1F1} -26 & {\cellcolor[HTML]{00285B}} \color[HTML]{F1F1F1} {\cellcolor[HTML]{00285B}} \color[HTML]{F1F1F1} -26 & {\cellcolor[HTML]{00285B}} \color[HTML]{F1F1F1} {\cellcolor[HTML]{00285B}} \color[HTML]{F1F1F1} -26 & {\cellcolor[HTML]{39486C}} \color[HTML]{F1F1F1} {\cellcolor[HTML]{39486C}} \color[HTML]{F1F1F1} -27 & {\cellcolor[HTML]{002758}} \color[HTML]{F1F1F1} {\cellcolor[HTML]{002758}} \color[HTML]{F1F1F1} -26 & {\cellcolor[HTML]{00224E}} \color[HTML]{F1F1F1} {\cellcolor[HTML]{00224E}} \color[HTML]{F1F1F1} -26 & {\cellcolor[HTML]{2E416D}} \color[HTML]{F1F1F1} {\cellcolor[HTML]{2E416D}} \color[HTML]{F1F1F1} -27 & {\cellcolor[HTML]{5E636F}} \color[HTML]{F1F1F1} {\cellcolor[HTML]{5E636F}} \color[HTML]{F1F1F1} -27 \\
2012-a & {\cellcolor[HTML]{00224E}} \color[HTML]{F1F1F1} {\cellcolor[HTML]{00224E}} \color[HTML]{F1F1F1} -156 & {\cellcolor[HTML]{A69C74}} \color[HTML]{F1F1F1} {\cellcolor[HTML]{A69C74}} \color[HTML]{F1F1F1} -172 & {\cellcolor[HTML]{BCAE6C}} \color[HTML]{000000} {\cellcolor[HTML]{BCAE6C}} \color[HTML]{000000} -174 & {\cellcolor[HTML]{FEE838}} \color[HTML]{000000} {\cellcolor[HTML]{FEE838}} \color[HTML]{000000} -180 & {\cellcolor[HTML]{747475}} \color[HTML]{F1F1F1} {\cellcolor[HTML]{747475}} \color[HTML]{F1F1F1} -167 & {\cellcolor[HTML]{6F7073}} \color[HTML]{F1F1F1} {\cellcolor[HTML]{6F7073}} \color[HTML]{F1F1F1} -167 & {\cellcolor[HTML]{8E8978}} \color[HTML]{F1F1F1} {\cellcolor[HTML]{8E8978}} \color[HTML]{F1F1F1} -170 & {\cellcolor[HTML]{C2B369}} \color[HTML]{000000} {\cellcolor[HTML]{C2B369}} \color[HTML]{000000} -175 & {\cellcolor[HTML]{32436D}} \color[HTML]{F1F1F1} {\cellcolor[HTML]{32436D}} \color[HTML]{F1F1F1} -161 & {\cellcolor[HTML]{203A6F}} \color[HTML]{F1F1F1} {\cellcolor[HTML]{203A6F}} \color[HTML]{F1F1F1} -159 & {\cellcolor[HTML]{233C6E}} \color[HTML]{F1F1F1} {\cellcolor[HTML]{233C6E}} \color[HTML]{F1F1F1} -160 & {\cellcolor[HTML]{50576C}} \color[HTML]{F1F1F1} {\cellcolor[HTML]{50576C}} \color[HTML]{F1F1F1} -163 \\
2012-b & {\cellcolor[HTML]{575D6D}} \color[HTML]{F1F1F1} {\cellcolor[HTML]{575D6D}} \color[HTML]{F1F1F1} -137 & {\cellcolor[HTML]{B1A570}} \color[HTML]{F1F1F1} {\cellcolor[HTML]{B1A570}} \color[HTML]{F1F1F1} -143 & {\cellcolor[HTML]{9C9576}} \color[HTML]{F1F1F1} {\cellcolor[HTML]{9C9576}} \color[HTML]{F1F1F1} -142 & {\cellcolor[HTML]{7F7E78}} \color[HTML]{F1F1F1} {\cellcolor[HTML]{7F7E78}} \color[HTML]{F1F1F1} -140 & {\cellcolor[HTML]{00224E}} \color[HTML]{F1F1F1} {\cellcolor[HTML]{00224E}} \color[HTML]{F1F1F1} -132 & {\cellcolor[HTML]{1E3A6F}} \color[HTML]{F1F1F1} {\cellcolor[HTML]{1E3A6F}} \color[HTML]{F1F1F1} -134 & {\cellcolor[HTML]{3C4A6C}} \color[HTML]{F1F1F1} {\cellcolor[HTML]{3C4A6C}} \color[HTML]{F1F1F1} -136 & {\cellcolor[HTML]{FEE838}} \color[HTML]{000000} {\cellcolor[HTML]{FEE838}} \color[HTML]{000000} -148 & {\cellcolor[HTML]{7D7C78}} \color[HTML]{F1F1F1} {\cellcolor[HTML]{7D7C78}} \color[HTML]{F1F1F1} -140 & {\cellcolor[HTML]{6F7073}} \color[HTML]{F1F1F1} {\cellcolor[HTML]{6F7073}} \color[HTML]{F1F1F1} -139 & {\cellcolor[HTML]{727374}} \color[HTML]{F1F1F1} {\cellcolor[HTML]{727374}} \color[HTML]{F1F1F1} -139 & {\cellcolor[HTML]{AEA371}} \color[HTML]{F1F1F1} {\cellcolor[HTML]{AEA371}} \color[HTML]{F1F1F1} -143 \\
2016~{\textcircled{{1}}} & {\cellcolor[HTML]{5B606E}} \color[HTML]{F1F1F1} {\cellcolor[HTML]{5B606E}} \color[HTML]{F1F1F1} -220 & {\cellcolor[HTML]{9E9676}} \color[HTML]{F1F1F1} {\cellcolor[HTML]{9E9676}} \color[HTML]{F1F1F1} -230 & {\cellcolor[HTML]{FEE838}} \color[HTML]{000000} {\cellcolor[HTML]{FEE838}} \color[HTML]{000000} -243 & {\cellcolor[HTML]{C3B369}} \color[HTML]{000000} {\cellcolor[HTML]{C3B369}} \color[HTML]{000000} -235 & {\cellcolor[HTML]{002B62}} \color[HTML]{F1F1F1} {\cellcolor[HTML]{002B62}} \color[HTML]{F1F1F1} -210 & {\cellcolor[HTML]{0C3470}} \color[HTML]{F1F1F1} {\cellcolor[HTML]{0C3470}} \color[HTML]{F1F1F1} -212 & {\cellcolor[HTML]{3E4B6C}} \color[HTML]{F1F1F1} {\cellcolor[HTML]{3E4B6C}} \color[HTML]{F1F1F1} -216 & {\cellcolor[HTML]{878478}} \color[HTML]{F1F1F1} {\cellcolor[HTML]{878478}} \color[HTML]{F1F1F1} -227 & {\cellcolor[HTML]{00224E}} \color[HTML]{F1F1F1} {\cellcolor[HTML]{00224E}} \color[HTML]{F1F1F1} -208 & {\cellcolor[HTML]{143670}} \color[HTML]{F1F1F1} {\cellcolor[HTML]{143670}} \color[HTML]{F1F1F1} -212 & {\cellcolor[HTML]{555B6D}} \color[HTML]{F1F1F1} {\cellcolor[HTML]{555B6D}} \color[HTML]{F1F1F1} -219 & {\cellcolor[HTML]{C1B26A}} \color[HTML]{000000} {\cellcolor[HTML]{C1B26A}} \color[HTML]{000000} -235 \\
2016~{\textcircled{{2}}} & {\cellcolor[HTML]{FEE838}} \color[HTML]{000000} {\cellcolor[HTML]{FEE838}} \color[HTML]{000000} -238 & {\cellcolor[HTML]{9D9576}} \color[HTML]{F1F1F1} {\cellcolor[HTML]{9D9576}} \color[HTML]{F1F1F1} -233 & {\cellcolor[HTML]{FEE838}} \color[HTML]{000000} {\cellcolor[HTML]{FEE838}} \color[HTML]{000000} -238 & {\cellcolor[HTML]{AAA073}} \color[HTML]{F1F1F1} {\cellcolor[HTML]{AAA073}} \color[HTML]{F1F1F1} -234 & {\cellcolor[HTML]{1A386F}} \color[HTML]{F1F1F1} {\cellcolor[HTML]{1A386F}} \color[HTML]{F1F1F1} -226 & {\cellcolor[HTML]{00224E}} \color[HTML]{F1F1F1} {\cellcolor[HTML]{00224E}} \color[HTML]{F1F1F1} -224 & {\cellcolor[HTML]{002859}} \color[HTML]{F1F1F1} {\cellcolor[HTML]{002859}} \color[HTML]{F1F1F1} -225 & {\cellcolor[HTML]{404C6C}} \color[HTML]{F1F1F1} {\cellcolor[HTML]{404C6C}} \color[HTML]{F1F1F1} -228 & {\cellcolor[HTML]{838179}} \color[HTML]{F1F1F1} {\cellcolor[HTML]{838179}} \color[HTML]{F1F1F1} -231 & {\cellcolor[HTML]{646770}} \color[HTML]{F1F1F1} {\cellcolor[HTML]{646770}} \color[HTML]{F1F1F1} -230 & {\cellcolor[HTML]{404C6C}} \color[HTML]{F1F1F1} {\cellcolor[HTML]{404C6C}} \color[HTML]{F1F1F1} -228 & {\cellcolor[HTML]{00295D}} \color[HTML]{F1F1F1} {\cellcolor[HTML]{00295D}} \color[HTML]{F1F1F1} -225 \\
2016~{\textcircled{{3}}} & {\cellcolor[HTML]{4C556C}} \color[HTML]{F1F1F1} {\cellcolor[HTML]{4C556C}} \color[HTML]{F1F1F1} -219 & {\cellcolor[HTML]{CEBC63}} \color[HTML]{000000} {\cellcolor[HTML]{CEBC63}} \color[HTML]{000000} -237 & {\cellcolor[HTML]{FEE838}} \color[HTML]{000000} {\cellcolor[HTML]{FEE838}} \color[HTML]{000000} -243 & {\cellcolor[HTML]{D0BE62}} \color[HTML]{000000} {\cellcolor[HTML]{D0BE62}} \color[HTML]{000000} -237 & {\cellcolor[HTML]{32436D}} \color[HTML]{F1F1F1} {\cellcolor[HTML]{32436D}} \color[HTML]{F1F1F1} -215 & {\cellcolor[HTML]{38476C}} \color[HTML]{F1F1F1} {\cellcolor[HTML]{38476C}} \color[HTML]{F1F1F1} -216 & {\cellcolor[HTML]{545A6D}} \color[HTML]{F1F1F1} {\cellcolor[HTML]{545A6D}} \color[HTML]{F1F1F1} -220 & {\cellcolor[HTML]{908B78}} \color[HTML]{F1F1F1} {\cellcolor[HTML]{908B78}} \color[HTML]{F1F1F1} -229 & {\cellcolor[HTML]{00224E}} \color[HTML]{F1F1F1} {\cellcolor[HTML]{00224E}} \color[HTML]{F1F1F1} -209 & {\cellcolor[HTML]{00234F}} \color[HTML]{F1F1F1} {\cellcolor[HTML]{00234F}} \color[HTML]{F1F1F1} -209 & {\cellcolor[HTML]{002C64}} \color[HTML]{F1F1F1} {\cellcolor[HTML]{002C64}} \color[HTML]{F1F1F1} -211 & {\cellcolor[HTML]{45506C}} \color[HTML]{F1F1F1} {\cellcolor[HTML]{45506C}} \color[HTML]{F1F1F1} -218 \\
2016~{\textcircled{{4}}} & {\cellcolor[HTML]{FEE838}} \color[HTML]{000000} {\cellcolor[HTML]{FEE838}} \color[HTML]{000000} -246 & {\cellcolor[HTML]{B8AA6E}} \color[HTML]{F1F1F1} {\cellcolor[HTML]{B8AA6E}} \color[HTML]{F1F1F1} -243 & {\cellcolor[HTML]{DEC958}} \color[HTML]{000000} {\cellcolor[HTML]{DEC958}} \color[HTML]{000000} -244 & {\cellcolor[HTML]{B5A86F}} \color[HTML]{F1F1F1} {\cellcolor[HTML]{B5A86F}} \color[HTML]{F1F1F1} -243 & {\cellcolor[HTML]{002E6C}} \color[HTML]{F1F1F1} {\cellcolor[HTML]{002E6C}} \color[HTML]{F1F1F1} -236 & {\cellcolor[HTML]{002859}} \color[HTML]{F1F1F1} {\cellcolor[HTML]{002859}} \color[HTML]{F1F1F1} -235 & {\cellcolor[HTML]{083370}} \color[HTML]{F1F1F1} {\cellcolor[HTML]{083370}} \color[HTML]{F1F1F1} -236 & {\cellcolor[HTML]{5D616E}} \color[HTML]{F1F1F1} {\cellcolor[HTML]{5D616E}} \color[HTML]{F1F1F1} -239 & {\cellcolor[HTML]{3A486C}} \color[HTML]{F1F1F1} {\cellcolor[HTML]{3A486C}} \color[HTML]{F1F1F1} -237 & {\cellcolor[HTML]{263D6E}} \color[HTML]{F1F1F1} {\cellcolor[HTML]{263D6E}} \color[HTML]{F1F1F1} -237 & {\cellcolor[HTML]{002E6C}} \color[HTML]{F1F1F1} {\cellcolor[HTML]{002E6C}} \color[HTML]{F1F1F1} -236 & {\cellcolor[HTML]{00224E}} \color[HTML]{F1F1F1} {\cellcolor[HTML]{00224E}} \color[HTML]{F1F1F1} -235 \\
2016~{\textcircled{{5}}} & {\cellcolor[HTML]{A99F73}} \color[HTML]{F1F1F1} {\cellcolor[HTML]{A99F73}} \color[HTML]{F1F1F1} -265 & {\cellcolor[HTML]{E5CF52}} \color[HTML]{000000} {\cellcolor[HTML]{E5CF52}} \color[HTML]{000000} -271 & {\cellcolor[HTML]{FEE838}} \color[HTML]{000000} {\cellcolor[HTML]{FEE838}} \color[HTML]{000000} -274 & {\cellcolor[HTML]{D7C45C}} \color[HTML]{000000} {\cellcolor[HTML]{D7C45C}} \color[HTML]{000000} -270 & {\cellcolor[HTML]{46516C}} \color[HTML]{F1F1F1} {\cellcolor[HTML]{46516C}} \color[HTML]{F1F1F1} -254 & {\cellcolor[HTML]{3B496C}} \color[HTML]{F1F1F1} {\cellcolor[HTML]{3B496C}} \color[HTML]{F1F1F1} -253 & {\cellcolor[HTML]{545A6D}} \color[HTML]{F1F1F1} {\cellcolor[HTML]{545A6D}} \color[HTML]{F1F1F1} -255 & {\cellcolor[HTML]{928C78}} \color[HTML]{F1F1F1} {\cellcolor[HTML]{928C78}} \color[HTML]{F1F1F1} -262 & {\cellcolor[HTML]{002F6D}} \color[HTML]{F1F1F1} {\cellcolor[HTML]{002F6D}} \color[HTML]{F1F1F1} -248 & {\cellcolor[HTML]{002656}} \color[HTML]{F1F1F1} {\cellcolor[HTML]{002656}} \color[HTML]{F1F1F1} -247 & {\cellcolor[HTML]{00224E}} \color[HTML]{F1F1F1} {\cellcolor[HTML]{00224E}} \color[HTML]{F1F1F1} -247 & {\cellcolor[HTML]{00306F}} \color[HTML]{F1F1F1} {\cellcolor[HTML]{00306F}} \color[HTML]{F1F1F1} -249 \\
2016~{\textcircled{{6}}} & {\cellcolor[HTML]{C4B468}} \color[HTML]{000000} {\cellcolor[HTML]{C4B468}} \color[HTML]{000000} -244 & {\cellcolor[HTML]{DFCA57}} \color[HTML]{000000} {\cellcolor[HTML]{DFCA57}} \color[HTML]{000000} -245 & {\cellcolor[HTML]{FEE838}} \color[HTML]{000000} {\cellcolor[HTML]{FEE838}} \color[HTML]{000000} -247 & {\cellcolor[HTML]{E5CF52}} \color[HTML]{000000} {\cellcolor[HTML]{E5CF52}} \color[HTML]{000000} -245 & {\cellcolor[HTML]{053371}} \color[HTML]{F1F1F1} {\cellcolor[HTML]{053371}} \color[HTML]{F1F1F1} -235 & {\cellcolor[HTML]{123570}} \color[HTML]{F1F1F1} {\cellcolor[HTML]{123570}} \color[HTML]{F1F1F1} -235 & {\cellcolor[HTML]{404C6C}} \color[HTML]{F1F1F1} {\cellcolor[HTML]{404C6C}} \color[HTML]{F1F1F1} -237 & {\cellcolor[HTML]{969077}} \color[HTML]{F1F1F1} {\cellcolor[HTML]{969077}} \color[HTML]{F1F1F1} -242 & {\cellcolor[HTML]{002C66}} \color[HTML]{F1F1F1} {\cellcolor[HTML]{002C66}} \color[HTML]{F1F1F1} -234 & {\cellcolor[HTML]{00224E}} \color[HTML]{F1F1F1} {\cellcolor[HTML]{00224E}} \color[HTML]{F1F1F1} -234 & {\cellcolor[HTML]{00224E}} \color[HTML]{F1F1F1} {\cellcolor[HTML]{00224E}} \color[HTML]{F1F1F1} -234 & {\cellcolor[HTML]{1C396F}} \color[HTML]{F1F1F1} {\cellcolor[HTML]{1C396F}} \color[HTML]{F1F1F1} -235 \\
2016~{\textcircled{{7}}} & {\cellcolor[HTML]{E0CB56}} \color[HTML]{000000} {\cellcolor[HTML]{E0CB56}} \color[HTML]{000000} -138 & {\cellcolor[HTML]{FEE838}} \color[HTML]{000000} {\cellcolor[HTML]{FEE838}} \color[HTML]{000000} -139 & {\cellcolor[HTML]{DAC65B}} \color[HTML]{000000} {\cellcolor[HTML]{DAC65B}} \color[HTML]{000000} -138 & {\cellcolor[HTML]{ABA072}} \color[HTML]{F1F1F1} {\cellcolor[HTML]{ABA072}} \color[HTML]{F1F1F1} -136 & {\cellcolor[HTML]{52596D}} \color[HTML]{F1F1F1} {\cellcolor[HTML]{52596D}} \color[HTML]{F1F1F1} -132 & {\cellcolor[HTML]{34456C}} \color[HTML]{F1F1F1} {\cellcolor[HTML]{34456C}} \color[HTML]{F1F1F1} -131 & {\cellcolor[HTML]{444F6C}} \color[HTML]{F1F1F1} {\cellcolor[HTML]{444F6C}} \color[HTML]{F1F1F1} -131 & {\cellcolor[HTML]{666970}} \color[HTML]{F1F1F1} {\cellcolor[HTML]{666970}} \color[HTML]{F1F1F1} -133 & {\cellcolor[HTML]{2D416D}} \color[HTML]{F1F1F1} {\cellcolor[HTML]{2D416D}} \color[HTML]{F1F1F1} -131 & {\cellcolor[HTML]{002C64}} \color[HTML]{F1F1F1} {\cellcolor[HTML]{002C64}} \color[HTML]{F1F1F1} -129 & {\cellcolor[HTML]{00224E}} \color[HTML]{F1F1F1} {\cellcolor[HTML]{00224E}} \color[HTML]{F1F1F1} -129 & {\cellcolor[HTML]{002A61}} \color[HTML]{F1F1F1} {\cellcolor[HTML]{002A61}} \color[HTML]{F1F1F1} -129 \\
\hline
Sum & -2187 &  -2289 & -2360 & -2337 & -2137 & -2157 & -2188 & -2293 & -2106 & -2111 & -2129 & -2192\\\hline\hline
\end{tabular}
\end{table*}

\begin{figure*}
    \centering
    \begin{center} 
    \includegraphics[width=0.8\textwidth]{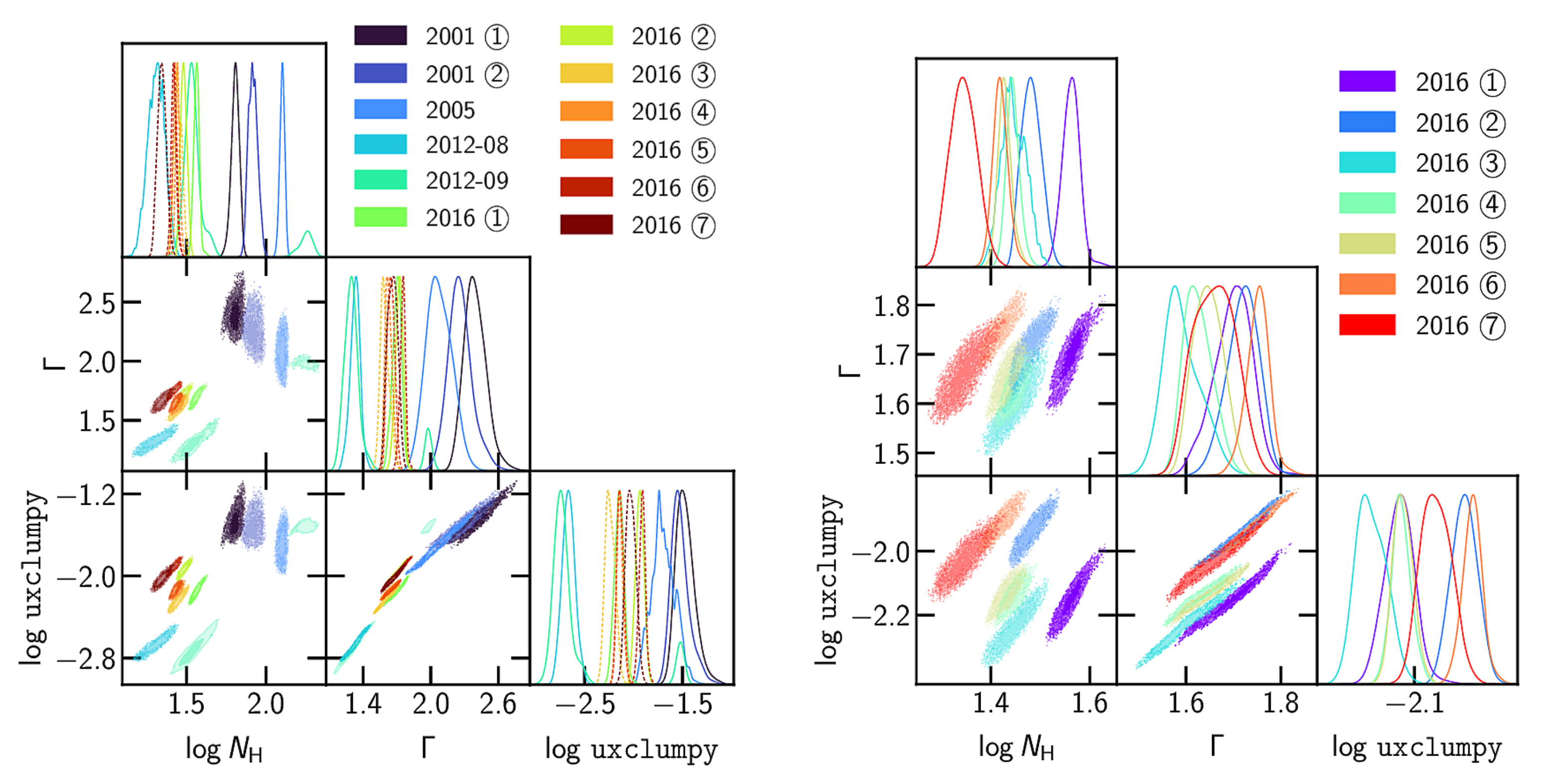}
 \end{center}
    \caption{Contours of the samples of the posterior distributions of the parameters of the physical model's best geometry. For all epochs (left) and all slices of the observation of 2016 (right).}
    \label{fig:physical_contours}
\end{figure*}

\begin{table*}
\caption{Historical values of the photon index $\Gamma$, the column density \NH\ and observed flux in the 2-10~keV band.$^\dag$ refers to parameters frozen or tied during the fits.}
\centering   \renewcommand{\arraystretch}{1.25}
\begin{tabular}{llllll}\hline\hline
Date & Observatory & $\Gamma$ &\begin{tabular}[c]{@{}l@{}} $N_\mathrm{H}$ \\$10^{22}~\mathrm{cm}^{-2}$\end{tabular} &\begin{tabular}[c]{@{}l@{}}Flux $(2-10)$~keV\\$10^{-12}$~\flux\end{tabular}& Reference  \\\hline
1984-06-09 & \textit{EXOSAT}&$1.70$ &$16.70_{-4.90}^{+5.80}$ &$17$ & \cite{1989MNRAS.240..833T}\\
1988-10-25 & \textit{Ginga}&$1.70$ &$48.40_{-15.20}^{+25.40}$ &$6$ & \cite{1993MNRAS.265..412W}\\
1994-11-14 & \textit{ASCA}&$1.38_{-0.11}^{+0.12}$ &$8.60_{-0.80}^{+0.80}$ &$15.10_{-4.20}^{+3.60}$&\cite{1997ApJS..113...23T} \\
1996-11-21 & \textit{ASCA}&$1.52_{-0.07}^{+0.09}$ &$12.40_{-0.80}^{+0.60}$ &$15.50_{-2.20}^{+2.40}$ &\cite{1998PASJ...50..519X}\\
1998-11-09 & \textit{BeppoSAX}&$1.95_{-0.18}^{+0.09}$ &$14.40_{-1}^{+0.90}$ &$19.70_{-3}^{+3}$&\cite{2000ApJ...531..245T}\\
2000-10-14 & \textit{Chandra} & $1.84_{-0.12}^{-0.12}$ &$17.95_{-2.11}^{+2.11}$ &$13.8$&\cite{2020ApJ...897...66L}\\
2000-10-15 & \textit{Chandra} &$1.45^\dag$ &$19.9_{-3.85}^{+3.85}$ &$14.1$&\cite{2020ApJ...897...66L}\\
2001-05-25 & \textit{XMM-Newton}&$1.90_{-0.02}^{+0.05}$ &$54.90_{-2.10}^{+6.60}$ &$4$ & \cite{2007AA...466..855P}\\
2003-06-01 & \textit{RXTE}&$2.10_{-0.10}^{+0.10}$ &-&$6.50_{-0.60}^{+0.60}$& \cite{2015ApJ...815...55R}\\
2003-10-01 & \textit{RXTE}&$2.10^{\dag}$ &$15_{-2}^{+2}$ &$13.20_{-0.60}^{+0.20}$ & \cite{2015ApJ...815...55R}\\
2004-02-01 & \textit{RXTE}&$2.10^\dag$&- &$6.50_{-0.20}^{+0.20}$ & \cite{2015ApJ...815...55R}\\
2004-08-01 & \textit{RXTE}&$2.10^\dag$ &$16_{-5}^{+5}$ &$8.80_{-0.20}^{+0.20}$ & \cite{2015ApJ...815...55R}\\
2005-04-29 & \textit{XMM-Newton}&$1.93_{-0.01}^{+0.01}$ &$129.30_{-6.70}^{+5.50}$ &$2.30$ & \cite{2007AA...466..855P}\\
2007-04-30 & \textit{XMM-Newton}&$1.92_{-0.16}^{+0.24}$ & $33_{-5}^{+4}$ &$7.60_{-0.40}^{+0.40}$ &\cite{2009ApJ...695..781B}\\
2007-05-01 & \textit{Suzaku}&$1.92^\dag$&$44_{-2}^{+3}$ &$5.30_{-0.30}^{+0.30}$&\cite{2009ApJ...695..781B} \\
2007-05-28 & \textit{Suzaku}&$1.92^\dag$ &$68_{-7}^{+6}$ &$4.10_{-0.30}^{+0.30}$ &\cite{2009ApJ...695..781B}\\
2007-11-09 & \textit{Suzaku}&$1.92^\dag$ &$110_{-11}^{+14}$ &$3.20_{-0.20}^{+0.20}$&\cite{2009ApJ...695..781B} \\
2007-11-16 & \textit{Suzaku}&$1.92^\dag$ &$120_{-20}^{+20}$ &$2.60_{-0.50}^{+0.50}$&\cite{2009ApJ...695..781B} \\
2012-08-31 & \textit{NuSTAR}&$1.78_{-0.07}^{+0.07}$ &$24_{-7}^{+3}$ &$4.8_{-0.1}^{+0.1}$& \cite{2015ApJ...815...55R} \\
2012-09-15 & \textit{NuSTAR}&$1.78_{-0.07}^{+0.07}$ &$56_{-20}^{+10}$&$3.2_{-0.1}^{+0.1}$ & \cite{2015ApJ...815...55R}\\
2014-11-16 & \textit{Chandra}&$1.74_{-0.18}^{+0.18}$& $120_{-20}^{+20}$ &$3.3$& \cite{2017AA...600A.135B}\\
2016-04-28 & \textit{XMM-Newton}&$1.44_{-0.02}^{+0.04}$ &$32_{-2}^{+1}$&$7.9$ & \cite{2022ApJS..260...30T}\\
2016-04-28 & \textit{NuSTAR}&$1.44^\dag$ &$32^\dag$&$7.9^\dag$ & \cite{2022ApJS..260...30T}\\\hline\hline
\end{tabular}
\label{tab:histoparams}
\end{table*}

\bsp	
\label{lastpage}
\end{document}